\documentclass[%
reprint,
 amsmath,amssymb,
 aps,
 prd,
]{revtex4-2}

\usepackage[colorlinks,allcolors=black]{hyperref}
\usepackage[nameinlink,capitalise]{cleveref}

\usepackage{graphicx}
\usepackage{dcolumn}
\usepackage{bm}
\usepackage{mathtools, stmaryrd}
\usepackage{xparse}
\usepackage[acronym]{glossaries}
\usepackage{printlen}
\usepackage{physics}
\usepackage{xcolor}

\newacronym[longplural=power spectral densities]{psd}{PSD}{power spectral density}
\newacronym[longplural=cross spectral densities]{csd}{CSD}{cross spectral density}
\newacronym{sgwb}{SGWB}{stochastic gravitational-wave background}
\newacronym{dft}{DFT}{discrete Fourier transform}
\newacronym{pdf}{PDF}{probability density function}
\newacronym{svd}{SVD}{singular value decomposition}
\newacronym{snr}{SNR}{signal-to-noise ratio}
\newacronym{tdi}{TDI}{time-delay interferometry}
\newacronym[longplural=degrees of freedom]{dof}{DOF}{degree of freedom}
\newacronym{rmse}{RMSE}{root mean squared error}
\newacronym{gw}{GW}{gravitational wave}
\newacronym{ldc}{LDC}{LISA Data Challenges}
\newacronym{lisa}{LISA}{Laser Interferometer Space Antenna}
\newacronym{oms}{OMS}{optical metrology system}
\newacronym{tm}{TM}{test mass}
\newacronym{pta}{PTA}{pulsar timing array}
\newacronym{wosa}{WOSA}{weighted overlapped segment averaging}

\Crefname{section}{Sec.}{Secs.}

\begin{document}

\preprint{APS/123-PRD}

\message{Textwidth is \printlength\columnwidth}

\title{Statistics of time and frequency-averaged spectra in gravitational-wave background searches}

\author{Quentin Baghi}
 \affiliation{Universit\'{e} Paris Cit\'{e}, CNRS, Astroparticule et Cosmologie, F-75013 Paris, France}
\author{Nikolaos Karnesis}%
\affiliation{%
Department of Physics, Aristotle University of Thessaloniki \\ Thessaloniki 54124, Greece \\
Research Center for Astronomy and Applied Mathematics, \\ Academy of Athens, Soranou Efessiou 4, \\ 11527 Athens, Greece
}%

\author{Jean-Baptiste Bayle}
\affiliation{%
IRFU, CEA, Universit\'{e} Paris-Saclay, F-91191 Gif-sur-Yvette, France
}%

\date{\today}

\begin{abstract}
Time series analysis from gravitational-wave detectors often relies on the assumption that time chunks, or frequency bins, are uncorrelated. We discuss the validity of this approximation in the context of searches for stochastic gravitational-wave backgrounds. We examine the impact of averaging over time and frequency, a reduction technique commonly employed to minimize the computational expense of likelihood evaluations. We introduce an analytical tool based on Fisher information to quantify the error in parameter inference arising from ignoring these effects. Finally, we address the issue of locally stationary processes and optimal time chunking. As an application, we study examples based on simulated data of the LISA space mission.
\end{abstract}

\maketitle


\section{Introduction}

Searching for stochastic signals in \gls{gw} detector data is a delicate task, as the signal shares common characteristics with the instrumental noise~\cite{christensen_stochastic_2019,romanoDetectionMethodsStochastic2017}. They can usually be described by random processes exhibiting some degree of stationarity and Gaussianity, as well as continuous spectra. Additionally, \gls{sgwb} searches require the analysis of long time series on a wide frequency range to make faint signals emerge, which requires a large volume of raw data.
To reduce the computational cost of such an analysis, it is necessary to choose a representation in which the noise covariance is sparse; this is typically achieved in the Fourier domain or using some time-frequency basis functions.
To further reduce the cost, a practical approach is to bin the data, i.e., chunk it in time or frequency and average spectral estimates. This is a common practice in spectral analysis, particularly in ground-based detector searches. The periodograms may be computed on many segments and then stacked~\cite{ligo2015,abbott_search_2019}, keeping each segment long enough to explore the lower end of the band. This estimation method is referred to as \gls{wosa}~\cite{nuttallSpectralEstimationUsing1982} or Welch's spectral estimation~\cite{welch_use_1967}.

While it is used in analyses requiring time chunking, the data compression obtained by averaging periodograms of time segments is limited by the lowest frequency of the band of interest. An alternative is to smooth spectral ordinates in the frequency domain; for example, by applying a moving average. This method has been used to reduce the frequency resolution in LIGO–Virgo analyses (e.g., \cite{abbott_upper_2017}) and in some recent studies relative to the future \gls{lisa} mission~\cite{caprini_reconstructing_2019,flauger_improved_2020,baghi_uncovering_2023,santini2025,buscicchioFirstYearLISA2025}.

Nevertheless, transforming data from the time domain to the frequency domain comes with spectral artifacts, which reflect in any subsequent averaging.
Any finite windowing (or tapering) leads to spectral leakage and correlations among frequency bins. These correlations affect the statistics of spectral estimates constructed from averaging periodograms. In Welch's approach, correlations of nearby segments decrease the effective number of averaged segments. In smoothing spectral estimates, a similar effect occurs from correlations between nearby frequency bins. Although sometimes mentioned~\cite{karnesis_assessing_2020}, these effects are not always accounted for in \gls{sgwb} searches, with some exceptions like in \cite{hindmarsh_recovering_2024}, where the authors account for the windowing in Welch's periodogram statistics. 

Spectral correlations have recently aroused interest in the \gls{gw} community.
\citet{sala2025precisionspectralestimationsubhz} have proposed a Bayesian approach to multivariate spectral estimation, specifically addressing correlations between data channels. They considered that frequency bins were sufficiently far apart to be treated as independent, as prescribed in \cite{vitaleDataSeriesSubtraction2014,armanoIndepthAnalysisLISA2024}. In the context of Advanced LIGO/Virgo, \citet{talbotInferenceFiniteTime2021} have recently investigated the effect of windowing and demonstrated that it can lead to errors in compact binary inference which are comparable to other known systematic effects like calibration errors. 
This type of investigation is also relevant to \gls{lisa} data analysis. While many studies in this context have been based on data directly simulated in the frequency domain, as LISA has passed the ``adoption'' milestone~\cite{lisa_adoption}, the tendency is to move towards more realistic assumptions and time-domain simulations. This is particularly the case for works that consider the search for anisotropic \glspl{sgwb} for their analysis, like \citet{adams_detecting_2014}, \citet{boileau_spectral_2021}, \citet{hindmarsh_recovering_2024}, \citet{criswell_templated_2024,criswellFlexibleSpectralSeparation2025},  \citet{buscicchioTestLISAForeground2025,pozzoliCyclostationarySignalsLISA2025,buscicchioFirstYearLISA2025}.
Besides, \citet{burke_mind_2025} have shown that the structure of the \gls{dft} covariance can significantly deviate from the diagonal assumption in the presence of data gaps, depending on the assumptions about the \gls{lisa} noise power spectrum. \citet{cornish2025nonstationary} has also investigated the conditions for having a diagonal covariance in time-frequency representations like wavelet packet bases. Even without interruptions, finite duration effects will likely have a significant impact on \gls{lisa} data analysis, given the level of expected \glspl{snr} for some \gls{gw} sources. Note that analyses of \glspl{pta} differ, as they are usually performed directly in the time domain, by modeling the unevenly sampled times of arrival using Gaussian processes (see, for example, \cite{vigelandBayesianInferencePulsartiming2014}). However, it has been shown~\cite{Bernardo2026} that the diagonal covariance approximation in frequency-domain \gls{pta} analysis can lead to systematic errors.



The first goal of this work is to characterize the distribution of smoothed spectral estimates (i.e., averaging periodograms in the frequency domain) for their use in \gls{sgwb} searches.
In contrast to approaches selecting frequency bins spaced enough to be considered uncorrelated, we study a smoothing approach that preserves the \gls{snr} by using all the available data within a frequency band, while accounting for bin-to-bin correlations. 
We employ smoothing as a compression strategy, with the aim of obtaining results that are statistically equivalent to those obtained with raw datasets.
We recall the impact of smoothing on power spectral density mean and variance, and we show how it translates into parameter uncertainties and biases when using parametrized spectral templates. We evaluate this bias using the Fisher information matrix, applying the rationale developed in~\cite{cutler2007} to spectral templates instead of waveforms. The tool allows us to find the optimal trade-off between parameter accuracy and spectral data compression, a manifestation of the infamous bias-variance compromise. We apply it to simulated observations of \gls{sgwb} with \gls{lisa}.

Secondly, we also discuss the effect of accounting for wide-sense non-stationarity when the spectrum is slowly evolving in time. Using the same Fisher-based formalism, we show how to quantify the impact of the time resolution on parameter accuracies.

We start in~\cref{sec:covariance} by reviewing standard definitions and results describing the covariance of stationary time series. Then, in~\cref{sec:time-binning}, we discuss the effect of correlations on time-averaged spectral estimates (Welch's method), and in~\cref{sec:frequency-binning} we consider the impact of frequency correlations on spectral smoothing. In~\cref{sec:inference}, we explore the implications on spectral parameter estimation by deriving an estimate of the bias whenever using an incorrect spectrum template. We generalize this framework to multivariate time series and locally stationary processes. In~\cref{sec:application-sgwb} we apply this formalism to a numerical experiment simulating a \gls{sgwb} search in \gls{lisa} data. We finally discuss our results in~\cref{sec:discussion}.

\section{\label{sec:covariance}Covariance of stationary time series}

\subsection{\label{sec:covariance-structure}Time and frequency-domain covariances}

We consider a time series $\mathbf{x}$ of size $N$ described by a zero-mean, Gaussian, weak-sense stationary process of covariance $\mathbf{\Sigma}_{xx}$. We assume it is sampled at a frequency $f_s$ (equivalent to a sampling time $\tau_s$). Hence, its duration is $T = N \tau_s$. 
Weak-sense stationarity implies that the covariance of two samples only depends on the duration separating them. Formally, this means that there exists an autocovariance function $R_{xx}(t)$ such that
\begin{equation}
    \Sigma_{xx}(i, j) = R_{xx}\left((j-i)\tau_s\right) \qcomma \forall i, j \in [0,\, N].
\end{equation}
The autocovariance function is related to the one-sided \gls{psd} function $S_{xx}(f)$ through an inverse Fourier transform:
\begin{equation}
    R_{xx}(t) = \frac{1}{2} \int_{-\frac{f_s}{2}}^{+\frac{f_s}{2}} S_{xx}(f) e^{2 i \pi f t} \dd{f}.
\end{equation}

Most of the analysis in gravitational-wave science is done in the Fourier domain. We define the windowed \gls{dft} of any time series $\mathbf{x}$ with the following convention:
\begin{equation}
\label{eq:wdft}
\tilde{x}_{N}(f_k) = \tau_s \sum_{n=0}^{N-1} w_n x_n e^{-2 i \pi \frac{nk}{N}},
\end{equation}
which has the same dimension as the continuous Fourier transform. In this definition, the index $k$ corresponds to frequency $f_k = k/T$ if $k \leq \lfloor \frac{N-1}{2} \rfloor$ and $f_k = (k-N)/T$ if $k > \lfloor \frac{N-1}{2} \rfloor$.
The definition includes a taper window $\mathbf{w}$ to prevent leakage effects. However, the classical \gls{dft} assumes $w_n = 1$ for all $n$.

Let us consider the structure of the covariance matrix in the frequency domain $\mathbf{\tilde{\Sigma}}$. One can show that its elements are
\begin{align}
	\label{eq:frequency-covariance-elements} 
    \tilde{\Sigma}_N(k, l) & = \operatorname{E}\left[\tilde{x}_{N}(f_k)\tilde{x}_{N}^{\ast}(f_l)\right] \nonumber \\ 
    & = \frac{1}{2}\int_{-\frac{f_s}{2}}^{+\frac{f_s}{2}} \tilde{w}_N(f_k - f) \tilde{w}_N^{\ast}(f_l - f) S_{xx}(f) \dd{f}.
\end{align}
This expression encodes two effects arising when considering finite, windowed time series: i) power leakage and ii) correlations of neighboring frequency bins. The first effect induces a distortion of the theoretical \gls{psd} in the diagonal elements of the covariance. This is due to the convolution with the squared modulus of the window's \gls{dft}, which deviates from a Dirac function for finite $N$ and exhibits side lobes. The second effect results in non-zero off-diagonal terms in the frequency-domain covariance, arising from the width of the main lobe of $\tilde{w}_N(f)$. Usually, windows tailored to reduce leakage tend to increase inter-bin correlations~\cite{heinzelSpectrumSpectralDensity2002}.

If all window points are non-zero, then the windowed \gls{dft} follows a complex Gaussian distribution whose \gls{pdf} is given by \cref{eq:complex-gaussian-pdf}. However, most tapering windows smoothly drop to zero at the edges of the time series. The consequence is that the matrix $\tilde{\Sigma}$ is almost singular (see~\cite{talbotInferenceFiniteTime2021,burke_mind_2025}). Indeed, windowing amounts to multiplying the time-domain covariance matrix on both sides by a diagonal matrix whose diagonal elements $(n, n)$ are equal to $w_n$. Therefore, the condition number of the resulting matrix is large. As an inverse covariance is required to compute the likelihood, one can resort to a regularization through truncated \gls{svd}.

\subsection{Spectrum estimation with the periodogram}

In \gls{sgwb} searches, basic spectrum estimation relies on the windowed periodogram that we define as
\begin{equation}
\label{eq:periodogram}
    P_{N}(f_k) \equiv \frac{2}{\tau_s \kappa_N}| \tilde{x}_{N}(f_k)|^2,
\end{equation}
with the normalization by the tapering window power $\kappa_N = \sum_{n=0}^{N-1} w_n^2$.
This definition is consistent with a one-sided \gls{psd}. Hence, the expectation of the windowed periodogram over infinite noise realizations is approximately given by $S(f_k)$, if one ignores frequency leakage:
\begin{equation}
\label{eq:periodogram-expectation-approx}
   \operatorname{E}[ P_{N}(f_k) ] \approx S(f_k).
\end{equation}
Finite-size windowing induces deviations from this approximation, as mentioned in~\cref{sec:covariance-structure}. When $N$ tends to infinity, this quantity is approximately proportional to a chi-squared distribution:
\begin{equation}
     P_{N}(f_k) \approx  
     \begin{cases}
     \frac{1}{2} S(f_k) \chi^2_{2} & \text{for} \; 0 < |f_k| < \frac{f_s}{2} \\
     S(f_k) \chi^2_{1}  & \text{for} \; |f_k| = 0 \text{ or } |f_k| = \frac{f_s}{2}.
\end{cases}
\end{equation}

\gls{sgwb} searches require using long observations to accumulate \gls{snr}. A direct analysis of the frequency-domain data is computationally expensive as $N$ gets large. Instead, many authors use binning, i.e., averaging periodograms obtained from contiguous data segments, or direct averaging in the frequency domain. We explore the implications of this strategy in the following.

\section{\label{sec:frequency-binning}Spectral smoothing}

In this work, we focus on frequency-domain binning as a data compression technique. We favor it over Welch's method, or \gls{wosa}, for two reasons. First, \gls{wosa} methods offer a moderate compression gain. For a given bandwidth of interest, the maximum data reduction is limited by the lowest observable frequency (determined by the segment length).
Second, \gls{wosa} techniques may be impractical when the underlying process exhibits non-stationarities, as the different segments cannot be assumed as identically distributed.

Another way of compressing the data is to compute the periodogram of the full time series and average its frequency bins. The simplest way is to split the Fourier band into $J$ segments. Each segment has size $N_j$ and is centered around a frequency $f_j$. We can then average all the periodogram bins inside each segment as
\begin{equation}
	\label{eq:averaged-periodogram}
    \bar{P}_{N}(f_j) \equiv \frac{1}{N_j} \sum_{k=j-K_j}^{j+K_j} P_{N}(f_k),
\end{equation}
where we took odd segment sizes $N_j = 2 K_j+1$ to get the same number of frequency bins below and above the central frequency. Note that for the sake of generality, we allow the number of averaged Fourier frequencies to depend on the segment index $j$. In the following, we discuss the consequences of using $\bar{P}_{N}(f_j)$ on spectral precision and accuracy.

\subsection{Approximate distribution of spectral estimates}

We aim at studying the statistics of $\bar{P}_{N}$. Let us first consider its mean:
\begin{equation}
	\label{eq:averaged-periodogram-mean}
    \mathrm{E}\left[\bar{P}_{N}(f_j)\right] = \frac{1}{2 K_j+1} \sum_{k=j-K_j}^{j+K_j} \bar{S}_{N}(f_k),
\end{equation}
where $\bar{S}_{N}(f_k) = \operatorname{E}[P_{N}(f_k)]$ is the expectation of the periodogram bins which is directly proportional to the diagonal elements of the frequency-domain covariance in \cref{eq:frequency-covariance-elements}. In general, the power leakage may result in distortions of the \gls{psd}. However, we assume that the window $w$ is well chosen so that they are negligible. We can write $\bar{S}_{N}(f_k) \approx S(f_k)$. If we Taylor-expand the \gls{psd} around the central frequency $f_j$ of the segment, we find
\begin{equation}
	\label{eq:averaged-periodogram-mean-2}
    \mathrm{E}\left[\bar{P}_{N}(f_j)\right] \approx S(f_j) + \frac{1}{2N_j} \sum_{k=j-K_j}^{j+K_j} S''(f_j) (f_k-f_j)^2,
\end{equation}
where $S''$ refers to the second derivative of the \gls{psd}. The first-order terms do not appear as they average out in the sum.

As each periodogram bin $P_{N}(f_k)$ is proportional to a chi-squared distribution with two degrees of freedom, if they were uncorrelated, their average would be proportional to a chi-squared distribution with $2N_j$ degrees of freedom. More precisely, we would have $2 \bar{P}_{N}(f_j) / S(f_j) \sim \chi^2_{2 N_j}$, assuming the \gls{psd} is constant over the entire segment.
However, we saw in~\cref{sec:covariance-structure} that different frequency bins are usually correlated over a certain bandwidth depending on the window. One approach would be to only include minimally correlated bins~\cite{vitaleDataSeriesSubtraction2014}, at the expense of a loss in \gls{snr}. We choose to keep all bins but to account for their correlations. To this aim, we can postulate that $\bar{P}_N(f)$ can be approximated to be proportional to a random variable following a chi-squared distribution with some \glspl{dof} $\nu < 2N_j$ to be adjusted~\cite{percival_spectral_2020}: 
\begin{equation}
\label{eq:pdf-propto-chi2}
    \bar{P}_N(f) \sim a \chi^2_{\nu}.
\end{equation}
In order for this distribution to have the same mean and variance, we require $\nu$ to be~\cite{noauthor_psd_1991}
\begin{equation}
\label{eq:welch-effective-ndof}
    \nu = \frac{2 \bar{S}_N(f)^2}{\operatorname{Var}\left[\bar{P}_N(f)\right]},
\end{equation}
and $a$ to be
\begin{equation}
    a = \frac{\bar{S}_N(f)}{\nu}.
\end{equation}
Note that the formulation of \cref{eq:pdf-propto-chi2} is perfectly equivalent to saying that $\bar{P}_N(f)$ follows a gamma distribution $\operatorname{Gamma}(\alpha, \theta)$ with shape parameter $\alpha = \nu / 2$ and scale $\theta = 2a$.

We can compute the effective number of \glspl{dof} from the ratio between the expectation and the variance of the averaged periodogram. The expression for the variance has the form:
\begin{align}
\label{eq:smooth-variance-approx-1}
    \operatorname{Var}\left[\bar{P}_N(f_j)\right] & \approx \frac{\operatorname{Var}\left[P_L(f_j)\right]}{N_j} \Bigg( 1 + \frac{2}{N_j} \sum_{\substack{k < k' \\ j-K_j}}^{j+K_j} \sigma_{kk'} \Bigg),
\end{align}
where $\sigma_{kk'}$ is the periodogram correlation coefficient among different frequencies:
\begin{equation}
\sigma_{kk'}  \equiv 
 \frac{\operatorname{Cov}[P_N(f_k), P_N(f_{k'})]}{\sqrt{\operatorname{Var}[P_{N}(f_k)]\operatorname{Var}[P_{N}(f_{k'})]}}.
\end{equation}

Plugging Eq.~\eqref{eq:smooth-variance-approx-1} into Eq.~\eqref{eq:welch-effective-ndof}, we obtain an expression for the effective number of~\glspl{dof}:
\begin{equation}
\label{eq:effective-ndof-freq}
    \nu_j = \frac{\nu^{0}_{j}}{1+\frac{2}{N_j} \sum_{k < k'} \sigma_{kk'}},
\end{equation}
where $\nu^{0}_{j} = 2 N_j$ is the number of \glspl{dof} if there were no correlations.
Similarly to the correlations of the different segment periodograms, the correlations among the different frequencies depend only on the window at first order:
\begin{equation}
\label{eq:correlation-coeff-frequency}
    \sigma_{kk'} \approx \gamma(k-k') = \frac{1}{\kappa_N^2}\left| \sum_{n=0}^{N-1} w_n^2 e^{-\frac{2i \pi (k-k')}{N}}\right|^2.
\end{equation}
The effective number of \glspl{dof} can then be rewritten as
\begin{equation}
\label{eq:effective-ndof-freq-approx}
    \nu_j = \frac{\nu^{0}_{j}}{1+ 2 \sum_{k = 1}^{N_j-1} \left(1-\frac{k}{N_j}\right) \gamma(k)}.
\end{equation}

We include the derivation of this result in \cref{sec:frequency-correlations}. Note that in the absence of correlations, $\gamma(k) = \delta(k)$, the number of \glspl{dof} is only given by the segment sizes $N_j$.
Note that a similar path can be followed to derive the effective number of \glspl{dof} for Welch's periodogram statistics, as we recall in~\cref{sec:time-binning}.

\subsection{\label{sec:example-frequency-averaging}Example on simulated samples}

We illustrate the effect of correlations in frequency binning with a numerical experiment. We consider a time series generated from a \gls{psd} $S_{xx}(f)$ describing the instrumental noise in the second-generation Michelson \gls{tdi} $X$ in the \gls{lisa} space mission (see, e.g., \cite{quang_nam_time-delay_2023} for a definition). The analytical expression for the \gls{psd} function is given in~\cref{sec:lisa-psd}. We choose $N = 8192$, $f_s = 6 \times 10^{-2}$ Hz, and generate $10^4$ realizations of the noise process. 

\begin{figure}
	 \includegraphics[width=\columnwidth]{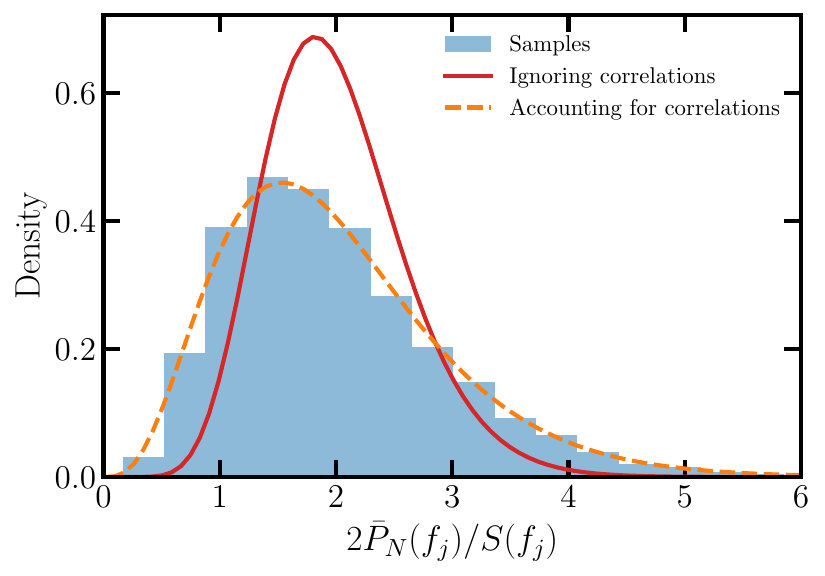}
	 \caption{\label{fig:averaged-periodogram}Averaged periodogram distribution using $10^4$ realizations of the same noise PSD for $f_j = 1$ mHz.}
 \end{figure}

We choose a frequency segment of size $N_j = 11$ centered around $f_j = 1$ mHz. For each realization, we compute the averaged periodogram following~\cref{eq:averaged-periodogram}, using a Blackman-Harris window, and the quantity $2 \bar{P}_{N}(f_j) / S_{N}(f_j)$. We obtain the histogram in~\cref{fig:averaged-periodogram} (in light blue). Naively trying to describe the distribution with a chi-squared \gls{pdf} with $2 N_j$ degrees of freedom leads to the red solid curve. We observe that this modeling is not accurate, largely overestimating the width of the distribution.

\begin{figure}
 	\includegraphics[width=\columnwidth]{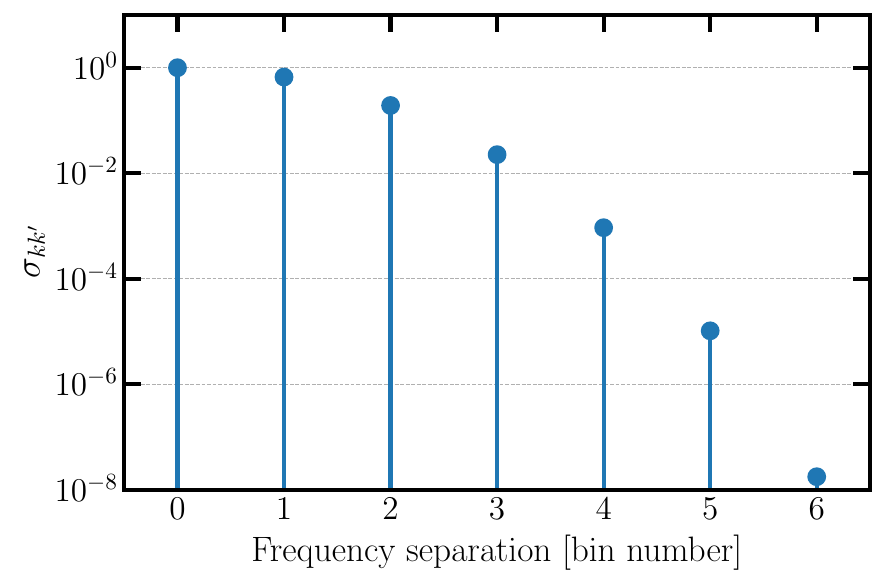}
 	\caption{\label{fig:correlation-vs-freq}Periodogram correlation coefficients $\sigma_{kk'}$ as a function of the frequency separation $k-k'$ (in number of bins) for the Blackman-Harris window with $N = 8192$.}
 \end{figure}

Then, we compute the theoretical correlation coefficients $\sigma_{kk'}$. For a fixed frequency bin $k$, we plot it as a function of the frequency bin separation in~\cref{fig:correlation-vs-freq}. The correlation is significant for bin separations up to 3, and then drops below $10^{-3}$ quickly. This is consistent with the conclusions drawn in~\cite{vitaleDataSeriesSubtraction2014} about the separation beyond which correlations can be ignored with this window. Therefore, one could approximate the correlation matrix by a banded matrix with a few non-zero diagonals.
 
 We compute the chi-squared distribution in \cref{eq:pdf-propto-chi2} using the effective number of \glspl{dof} computed with \cref{eq:effective-ndof-freq}. We obtain the dashed orange curve in~\cref{fig:averaged-periodogram}, which better agrees with the empirical distribution. We check this agreement by performing a Kolmogorov-Smirnov test, which yields a p-value of $0.27$. This is above the significance level $0.05$, in favor of the samples being distributed according to the analytical distribution.
 The ratio of the standard deviations of the correlation-aware and correlation-blind distributions is about $(\nu_j/\nu^{0}_{j})^{1/2} \sim 0.63$. Therefore, ignoring bin-to-bin correlations in this case would result in the uncertainties of inferred astrophysical or cosmological parameters being underestimated by~37\%.

\subsection{Comparison with frequency subsampling}

An alternative approach to the variance correction is to use the fact that the frequency ordinates can be considered uncorrelated above a bin separation $p$, and discard every other $p$ frequency samples, as in~\cite{vitaleDataSeriesSubtraction2014}. This way, each averaged estimate would be given by
\begin{equation}
	\label{eq:averaged-periodogram-subsampling}
    \bar{P}^{p}_{N}(f_j) \equiv \frac{p}{N_j} \sum_{n=-K_j/p}^{+K_j/p} P_{N}(f_{j+np}),
\end{equation}
and its variance would be
\begin{align}
	\label{eq:smooth-variance-approx-1-subsampling}
    \operatorname{Var}[\bar{P}^{p}_{N}(f_j)] & = \frac{p \operatorname{Var}\left[P_L(f_j)\right]}{N_j} \Bigg[
     1 + \frac{2}{N_j} \sum_{\substack{n < n' \\ (j-K_j)/p}}^{(j+K_j)/p} \sigma_{np, n'p} \Bigg]\nonumber\\
    & \approx \frac{p}{N_j} \operatorname{Var}[P_{L}(f_j)],
\end{align}
instead of \cref{eq:smooth-variance-approx-1}.
We can quantify the relative performance in terms of standard deviation (or \gls{snr}) of the two approaches by computing the uncertainty ratio
\begin{align}
\label{eq:full-vs-subsampled-variance-ratio}
        r_p & = \sqrt{\frac{\operatorname{Var}[\bar{P}^{p}_{N}(f_j)]}{\operatorname{Var}[\bar{P}_{N}(f_j)]}} \nonumber \\
        & \approx \sqrt\frac{p\left[1+ 2 \sum_{n = 1}^{N_j/p-1} \left(1-\frac{np}{N_j}\right) \gamma(np)\right]}{1+ 2 \sum_{k = 1}^{N_j-1} \left(1-\frac{k}{N_j}\right) \gamma(k)} \nonumber \\
        & \approx \sqrt{ \frac{p \, \nu_j}{\nu^{0}_{j} }},
\end{align}
where in the second line we used the fact that the correlations weakly depend on the \gls{psd} and can be expressed with the correlation coefficient $\gamma(k)$ depending only on the window and the bin difference. In the last line, we assumed that all correlations with separations larger than or equal to $p$ can be discarded, so that the numerator reduces to $p$.
The result in \cref{eq:full-vs-subsampled-variance-ratio} can be interpreted in the following way: using the full average rather than subsampling is more advantageous every time $r_p > 1$, which approximately gives $p > \nu_j^{0}/\nu_j$. In the example of \cref{sec:example-frequency-averaging}, we saw that correlations could be considered negligible beyond a separation of 3 bins, suggesting we can retain every $p=4$ bins. We evaluate $\nu_j^{0}/\nu_j \sim 2.6$, which yields a standard deviation ratio of $r_4 \sim 1.25$, suggesting that the full average is preferable to the subsampled version.

Keeping the same segment size $N_j$, we compute $r_p$ for a few different windows in \cref{tab:variance-ratios}, assuming that correlations can be safely ignored for $\gamma(k) \leq 10^{-3}$.
\begin{table}[h]
    \caption{Number of bins beyond which correlations are negligible and uncertainty ratio between the subsampled average and the full average periodograms, for a few different taper windows.}
    \label{tab:variance-ratios}
    \centering
    \begin{tabular}{lcc}
        \hline
        Window Type & Correlation length $p$ & Uncertainty Ratio $r_p$ \\
        \hline
        Blackman-Harris & 4 & 1.25 \\
        Hann     & 3 & 1.27 \\
        Hamming  & 3 & 1.32 \\
        Blackman & 4 & 1.35 \\
        Flattop  & 6 & 1.29 \\
        \hline
    \end{tabular}
\end{table}
All uncertainty ratios remain above unity in these settings. For a signal of constant \gls{psd} within a bin, the \gls{snr} ratio between the subsampled and the full average is proportional to $r_{p}^{-1}$; therefore, constructing periodogram averages over all frequency bins is always better for \gls{snr} preservation.

\subsection{Bias-variance trade-off}

Using \cref{eq:averaged-periodogram} to average periodogram bins in the frequency domain decreases the variance of spectral estimates while increasing the bias. The segment size $N_j$ controls this trade-off. Therefore, looking for the best data compression requires finding the maximum value of $N_j$ for which the bias remains low enough.

\subsubsection{For spectral estimates}

If one is interested in computing accurate spectral estimates with the minimal variance, the \gls{rmse} $\epsilon$ is a possible figure of merit. Let us consider a frequency bandwidth $B = f_{\max} - f_{\min}$. We define the \gls{rmse} $\epsilon$ over $B$ as
\begin{equation}
    \epsilon^2 = \frac{1}{N_B}\sum_{j=j_{\min}}^{j_{\max}} \left(\bar{P}_N(f_j) - S(f_j) \right)^2,
\end{equation}
where we choose the lower bound of the first frequency segment (respectively the upper bound of the last segment) to coincide with the bandwidth's lower bound (respectively upper bound): $f_{\min} = f_{j_{\min}} - K_j/T$; $f_{\max} = f_{j_{\max}} + K_j/T$. We also set the number of frequency bins in the bandwidth $N_B = j_{\max} - j_{\min}+1$.
Rather than the \gls{rmse}, which depends on the realization of the stochastic process, we consider the statistical expectation of $\epsilon^2$:
\begin{align}
    \operatorname{E}[\epsilon^2] & = \frac{1}{N_B}\sum_{j=j_{\min}}^{j_{\max}} \operatorname{Var}[\bar{P}_N(f_j)] + \operatorname{Bias}^2[\bar{P}_N(f_j)], \nonumber \\
    & = \sigma^2 + b^2
\end{align}
where the variance of the averaged spectra is approximately
\begin{equation}
    \operatorname{Var}[\bar{P}_N(f_j)] \approx \frac{2}{\nu(f_j)} S(f_j)^2,
\end{equation}
and we defined the bias as
\begin{equation}
\label{eq:spectrum-bias}
    \operatorname{Bias}[\bar{P}_N(f_j)] \equiv \operatorname{E}[\bar{P}_N(f_j)] - S(f_j).
\end{equation}
The quantities $\sigma^2$ and $b^2$ are the contributions of the total variance and of the total bias to the \gls{rmse}, respectively. 
From \cref{eq:averaged-periodogram-mean-2}, one can evaluate the bias as a function of the second derivatives of the \gls{psd}:
\begin{align}
    \operatorname{Bias}[\bar{P}_N(f_j)] & \approx \frac{1}{2N_j} \sum_{k=j-K_j}^{j+K_j} S''(f_j) (f_k-f_j)^2 \nonumber \\
    & = S''(f_j) \frac{K_j(K_j+1)}{6 T^2},
\end{align}
where in the last line we used the analytical formula for the sum of squared integers. As expected, the larger the \gls{psd} second derivative, the larger the bias. It also increases with the square of the averaging bandwidth $K_j/T$.

\begin{figure}
 	\includegraphics[width=\columnwidth]{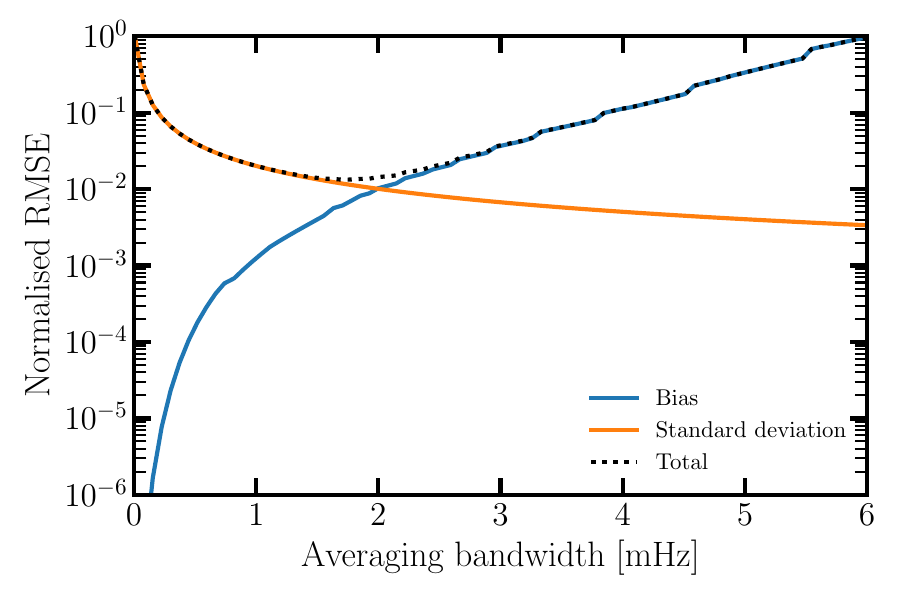}
 	\caption{\label{fig:spectrum-bias-variance}Expected bias and variance of smoothed spectral estimates (normalized by the \gls{psd}) as a function of the averaging bandwidth, for an observation time of about $1.4 \times 10^5$ seconds.}
 \end{figure}

To find an optimal trade-off between bias and variance, one can compute the total bias $b$ and the total standard deviation $\sigma$ normalized by the \gls{psd} as a function of the averaging bandwidth. We present the results in~\cref{fig:spectrum-bias-variance} for the same case study as previously. The plot indicates that the crossing point is located around 2 mHz. This means that above this bandwidth, the bias is significant compared to the periodogram variance. This computation offers a simple tool to tune smoothing in spectrum estimation, provided that an estimate of the \gls{psd} is available.

 However, when inferring parameters controlling the \gls{psd}, this type of assessment may not be accurate enough. In particular, if some parameters influence different frequency regions or faint features in the spectrum, the global trade-off provided by \cref{fig:spectrum-bias-variance} may underestimate the maximum averaging one can do without introducing a significant bias in parameter estimates. We address this issue in the following section.

\section{\label{sec:inference}Bias and variance in parameter inference}

So far, we have studied the effect of averaging on spectral estimates themselves.
In this section, we extend the study to the case where the \gls{psd} depends on unknown parameters that must be estimated.

\subsection{Gaussian stationary likelihood}

Performing parameter estimation in a classical Bayesian framework requires writing down a likelihood function. Under the assumptions stated in \cref{sec:covariance}, we can write the \gls{pdf} of the time series $\mathbf{x}$ as
\begin{equation}
\label{eq:pdf-time}
    f_{\mathbf{X}}\left(\mathbf{x}\right)=\frac{\exp \left(-\frac{1}{2}\mathbf{x}^{\mathrm{T}} \mathbf{\Sigma}^{-1}\mathbf{x}\right)}{\sqrt{(2 \pi)^N|\mathbf{\Sigma}|}}.
\end{equation}

For now, let us assume that $w_n \neq 0 \; \forall n$. Then, the windowed \gls{dft} vector $\mathbf{\tilde{x}}_{N}$ of entries $\tilde{x}_{N}(f_k)$ follows a zero-mean complex Gaussian distribution:
\begin{equation}
\label{eq:complex-gaussian-pdf}
      f_{\mathbf{\tilde{X}}_{N}} \left( \mathbf{\tilde{x}}_N \right) = \frac{\exp \left(-\mathbf{\tilde{x}}_N^{\dagger} \mathbf{\tilde{\Sigma}}_{N}^{-1}\mathbf{\tilde{x}}_N \right)}{\pi^N|\mathbf{\tilde{\Sigma}}_N|} ,
\end{equation}
where $\tilde{\Sigma}_N$ is the covariance of $\mathbf{\tilde{x}}_N$. One can show that $\tilde{\Sigma}_N$ is a Toeplitz matrix, which means it has constant diagonals. 

\subsection{Whittle's approximation of the likelihood}

Whittle's approximation consists of ignoring the exact structure of the covariance and assumes that $\tilde{\Sigma}_N$ is circulant, which is achieved when $N$ tends to infinity. Following this assumption, the frequency-domain covariance is approximately diagonal:
\begin{equation}
    \mathbf{\tilde{\Sigma}}_N \approx \operatorname{diag}(\boldsymbol{\Delta}),
\end{equation}
and the diagonal elements $\boldsymbol{\Delta}$ are proportional to the \gls{psd}, so that
\begin{equation}
\label{eq:diagonal-approx}
    \Delta(k) = \frac{\kappa_N \tau_s}{2} S(f_k).
\end{equation}
Therefore, Whittle's approximation has two aspects: by ignoring windowing effects, it disregards correlations among frequency bins (i.e., the off-diagonal terms of the covariance matrix) and the effects of leakage on neighboring frequencies (i.e., diagonal term distortions). Note, however, that leakage may be accounted for while preserving the same likelihood form~\cite{sykulskiDebiasedWhittleLikelihood2019,astfalck_debiasing_2024}.  \cref{eq:complex-gaussian-pdf} is approximated by
\begin{equation}
\label{eq:whittle-pdf}
      f_{\mathbf{\tilde{X}}_N} \left( \mathbf{\tilde{x}}_N \right) \approx \frac{\exp \left(- \sum_{k=0}^{N-1} \frac{|\tilde{x}_{N}(f_k)|^2}{\Delta(k)} \right)}{\pi^N \Pi_{k=0}^{N-1} \Delta(k)}.
\end{equation}
We call this \gls{pdf} Whittle's likelihood. When taking the logarithm, it takes the well-known form
\begin{equation}
    \log f_{\mathbf{\tilde{X}}_N} = - \sum_{k=0}^{N-1} \left[\frac{|\tilde{x}_{N}(f_k)|^2}{\Delta(k)}  + \log \Delta(k) \right]
\end{equation}
where we ignored the constant terms.

\subsection{Whittle's likelihood with binned data}

We saw that binning modifies the expectation-to-variance ratio of the distribution, which is directly related to the effective number of \glspl{dof}. Let us call $\bar{P}$ a spectral estimate obtained by averaging, either in time (Welch's method) or in frequency (spectral smoothing).
From the approximation in \cref{eq:pdf-propto-chi2}, one can derive the log-likelihood of averaged spectral estimates:
\begin{equation}
    \label{eq:averaged-loglike}
        \log f_{\bar{P}} = - \sum_{j=j_{\min}}^{j_{\max}} \frac{\nu_{j}}{2} \left[\frac{\bar{P}(f_j)}{\bar{S}(f_j)}  + \log \bar{S}(f_j) \right],
\end{equation}
where we dropped the constant terms and used $\bar{S} = \operatorname{E}[\bar{P}]$. We also restricted the analysis to some bandwidth set by $j_{\min}$ and $j_{\max}$. 
We labeled $\nu_j$ the effective number of \glspl{dof} accounting for the correlations among the averaged periodograms.

One can think of \cref{eq:averaged-loglike} as an extension of Whittle's likelihood, accounting for window or averaging distortions (using $\bar{S}$ instead of $S$) and correlations between frequencies within the segments (using $\nu$ instead of twice the number of averaged periodograms). Using $S$ instead of $\bar{S}$ may introduce bias in parameter estimates, whereas using the number of averaged periodograms instead of $\nu$ may lead to underestimated parameter uncertainties. Besides, this formulation does not account for correlations between \textit{averaged} periodograms at different central frequencies $j$.

For well-chosen windows and sufficiently long time series, the approximation $\bar{S} \approx S$ is acceptable. However, we see from \cref{fig:averaged-periodogram} that changes in the variance may be significant. In the case of a constant number of averaged periodograms $N_j = N_{\mathrm{seg}}$ across frequencies, any inference of parameters describing the \gls{psd} will yield standard deviations that are underestimated by a factor $\sqrt{\nu/(2N_{\mathrm{seg}})}$.

\subsection{Bias estimation}

The optimal bias-variance trade-off depends on the specific focus of the analysis. For example, if the aim is to estimate the parameters of a low-frequency component, the optimal binning will be coarser than if the aim is to estimate a parameter that drives the high-frequency end of the spectrum.

For any given parameter and an associated likelihood model, Fisher's information provides a forecast of parameter errors, or at least a lower bound. Let us assume that the \gls{psd} is parametrized by a vector of $p$ parameters $\boldsymbol{\theta}$, so that $S = S(f, \boldsymbol{\theta})$. The Fisher information matrix has entries
\begin{equation}
    I_{lm}(\boldsymbol{\theta}) = - \operatorname{E}\left[\frac{\partial^2 \mathcal{L}}{\partial \theta_l \partial \theta_m}\right],
\end{equation}
where $\mathcal{L}$ is the log-likelihood.
Using the extended Whittle's likelihood in \cref{eq:averaged-loglike} for $\mathcal{L} = \log f_{\bar{P}}$, and assuming $\bar{S}(f) \approx S(f)$, one obtains
\begin{equation}
\label{eq:fisher}
    I_{lm}(\boldsymbol{\theta}) =  \sum_{j=j_{\min}}^{j_{\max}} \frac{\nu_j}{2} \frac{1}{S(f_j)^2} \frac{\partial S(f_j)}{\partial \theta_l}  \frac{\partial S(f_j)}{\partial \theta_m}.
\end{equation}

In addition to forecasting statistical uncertainties, estimating the bias can be of key interest when choosing the characteristics of spectral estimates, such as the level of smoothing. We can proceed similarly to Cutler \& Vallisneri \cite{cutler2007}. In their work, the authors evaluate parameter errors when using inaccurate \gls{gw} waveform templates. For \gls{sgwb} searches, we rather focus on parametric \glspl{psd}. We introduce the best-fit parameter $\boldsymbol{\theta}_{\mathrm{bf}}$, i.e., the one that maximizes the log-likelihood $\mathcal{L}$. We have, by definition:
\begin{equation}
\label{eq:best-fit-equation}
 \boldsymbol{\nabla} \mathcal{L}(\boldsymbol{\theta}_\mathrm{bf}) = \mathbf{0},
\end{equation}
where $\boldsymbol{\nabla} \mathcal{L}$ denotes the log-likelihood gradient.

We label the actual parameter describing the analyzed data (the one that drives the stochastic process generation) as  $\boldsymbol{\theta}_{\mathrm{tr}}$.
We want to evaluate the difference $\Delta \boldsymbol{\theta} = \boldsymbol{\theta}_{\mathrm{bf}} - \boldsymbol{\theta}_{\mathrm{tr}}$, which is the error made when using the theoretical \gls{psd} $S(f)$ instead of $\bar{S}(f)$ in the likelihood, i.e., when ignoring the effects of averaging and windowing on the periodogram expectation. We show in \cref{sec:bias-proof} that an estimate of this error is 
\begin{equation}
\label{eq:parameter-bias}
     \Delta \boldsymbol{\theta} \approx \mathbf{I}(\boldsymbol{\theta}_{\mathrm{bf}})^{-1} \mathbf{B},
\end{equation}
where $\mathbf{B}$ is the $p-$column vector whose entries are
\begin{align}
    \label{eq:def-rhs-term}
    B_{l} \equiv \sum_{j=j_{\min}}^{j_{\max}} \frac{\nu_j}{2}\frac{\partial S(f_j)}{\partial \theta_l}  \frac{\operatorname{Bias}[\bar{P}(f_j) \, |\,  \boldsymbol{\theta}_{\mathrm{tr}}]}{S(f_j, \boldsymbol{\theta}_\mathrm{bf})^2},
\end{align}
which unsurprisingly depends on the spectrum bias introduced in \cref{eq:spectrum-bias}. Note that this result generalizes to any type of inaccuracies of the \gls{psd} template, including physical modeling errors. In addition, to leading order, $\boldsymbol{\theta}_\mathrm{tr}$ can be replaced by $\boldsymbol{\theta}_\mathrm{bf}$ in \cref{eq:def-rhs-term}. This is convenient when performing an a-posteriori assessment where the true parameter is unknown.

\subsection{\label{sec:multivariate}The multivariate case: covariance and spectral estimate}

Up to now, we have considered a single observable in the time series. This section addresses the case of the joint analysis of several time series representing different observables. This is essential in multi-detector ground-based searches for stochastic \glspl{gw}~\cite{romanoDetectionMethodsStochastic2017}, where the use of multiple measurements enables disentangling signal and noise. This will also be the case of third-generation detectors in triangular configuration, although noise-sourced correlations have to be addressed~\cite{caporaliImpactCorrelatedNoise2025}. \gls{lisa} faces the same challenge, where any complete set of variables necessary for science extraction exhibits cross-correlations due to both noise and signal. This is the case of the Michelson \gls{tdi} variables $X$, $Y$, $Z$~\cite{tinto_time-delay_2020}.  We consider them as three dependent, stationary, and Gaussian-distributed time series of length $N$, represented by three column vectors $\mathbf{x}$, $\mathbf{y}$, $\mathbf{z}$. We can form the full data vector of size $3N$ as $\mathbf{d} = \left(\mathbf{x}^{\intercal}, \, \mathbf{y}^{\intercal}, \, \mathbf{z}^{\intercal}\right)^{\intercal}$.

The frequency-domain covariance matrix of $\mathbf{\tilde{d}}$ can be partitioned into $N \times N$ blocks 
\begin{equation}
\label{eq:full-multivariate-cov}
	\mathbf{\tilde{\Sigma}}_{d} = \begin{pmatrix}
		\mathbf{\tilde{\Sigma}}_{xx} & \mathbf{\tilde{\Sigma}}_{xy} & \mathbf{\tilde{\Sigma}}_{xz} \\
		\mathbf{\tilde{\Sigma}}_{yx} & \mathbf{\tilde{\Sigma}}_{yy} & \mathbf{\tilde{\Sigma}}_{yz} \\
		\mathbf{\tilde{\Sigma}}_{zx} & \mathbf{\tilde{\Sigma}}_{zy} & \mathbf{\tilde{\Sigma}}_{zz} 
	\end{pmatrix}
\end{equation}
where
\begin{equation}
\mathbf{\tilde{\Sigma}}_{\alpha \beta} \equiv \operatorname{Cov}\left( \mathbf{\tilde{\alpha}}, \mathbf{\tilde{\beta}}  \right)  \; \; \forall \alpha, \beta \in \{x, y, z\}.
\end{equation}
We can write down an expression for the element of the block matrices by generalizing \cref{eq:frequency-covariance-elements}:
\begin{equation}
	\label{eq:frequency-covariance-elements-multivariate}
	\tilde{\Sigma}_{\alpha \beta}(i, j) = \frac{1}{2}\int_{-\frac{f_s}{2}}^{+\frac{f_s}{2}} \tilde{w}(f_i - f) \tilde{w}^{\ast}(f_j - f) S_{\alpha \beta}(f) df,
\end{equation}
where $S_{\alpha \beta}(f)$ designates the \gls{csd} function when $\alpha \neq \beta$.
This multivariate data can also be modeled as a Gaussian distribution of the same form as \cref{eq:complex-gaussian-pdf}.

We define the periodogram matrix as the $3 \times 3$ frequency-dependent matrix
\begin{equation}
\label{eq:periodogram-matrix}
    \mathbf{P}_{N}(f_k) \equiv \frac{2}{\tau_s \kappa_N}\mathbf{\tilde{d}}(f_k) \mathbf{\tilde{d}}(f_k)^\dagger,
\end{equation}
where $\mathbf{\tilde{d}}(f_k)$ is the $3 \times 1$ column vector representing the \gls{dft} of the \gls{tdi} channels evaluated at frequency bin $f_k$.

Like in the univariate case, we can consider short segments in the time or frequency domain. Without loss of generality, we focus on frequency-domain segments here.
Instead of averaging over scalar periodograms, we need to average over sample covariance matrices. This way, we define an averaged $3 \times 3$ periodogram matrix for each frequency segment:
\begin{equation}
\label{eq:multivariate-binning}
	\mathbf{\bar{P}}(f_j) \equiv \frac{1}{N_j} \sum_{k=j-K_j}^{j+K_j} \mathbf{P}_{N}(f_k).
\end{equation}
The individual sample covariances entering the sum are usually correlated due to finite windowing, like in the univariate case.

\subsection{\label{sec:multivariate-likelihood}Multivariate approximate likelihood}

We saw that in the univariate case, one can approximate the averaged periodogram as a random variable proportional to a chi-squared distributed variable. The generalization of this approximation in the multivariate case amounts to assuming that $\mathbf{\bar{P}}$ is proportional to a Wishart distribution with an effective number of degrees of freedom $n = \nu/2$. The \gls{pdf} then writes
\begin{equation}
    f_{\bar{P}}(\mathbf{\bar{P}})=\frac{|\mathbf{\bar{P}}|^{n-3} e^{-n \operatorname{tr}\left(\mathbf{C}^{-1} \mathbf{\bar{P}}\right)}}{|n^{-1}\mathbf{C}|^n \cdot \mathcal{C} \widetilde{\Gamma}_p(n)},
\end{equation}
where $\mathcal{C}\widetilde{\Gamma}_p(n)$ is the complex multivariate Gamma function.
After dropping the constant terms, one can write the log-\gls{pdf} of $\mathbf{\bar{P}}$ as
\begin{equation}
\label{eq:wishart_loglike}
    \log f_{\mathbf{\bar{P}}} = - \sum_{j=j_{\min}}^{j_{\max}} \frac{\nu_j}{2} \left(\log |\mathbf{C}(f_j)| + \operatorname{tr}\left(\mathbf{C}(f_j)^{-1} \mathbf{\bar{P}}(f_j) \right)\right)
\end{equation}
where the frequency-dependent matrix $\mathbf{C}(f)$ is Hermitian and has dimension $3 \times 3$ with elements
$\mathbf{C}_{\alpha\beta}(f) = S_{\alpha\beta}(f)$ for $\alpha \leq \beta$. Note that this formulation ignores windowing effects, as we assume that the covariance blocks $\boldsymbol{\tilde{\Sigma}_{\alpha\beta}}$ are diagonal and their diagonal elements are assimilated to the \glspl{psd} and \glspl{csd}.

\subsection{\label{sec:multivariate-bias}Multivariate bias estimate}

The multivariate version of the Fisher matrix in \cref{eq:fisher}  can be written as
\begin{equation}
\label{eq:multivariate-fisher}
    I_{lm}(\boldsymbol{\theta}) =  \sum_{j=j_{\min}}^{j_{\max}} \frac{\nu_j}{2} \operatorname{tr}\left(\mathbf{C}(f_j)^{-1}\frac{\partial \mathbf{C}(f_j)}{\partial \theta_l}\mathbf{C}(f_j)^{-1}  \frac{\partial \mathbf{C}(f_j)}{\partial \theta_m}\right).
\end{equation}
Equation \eqref{eq:parameter-bias} for the parameter bias remains the same, where the elements of vector $\mathbf{B}$ are now given by
\begin{align}
\label{eq:multivariate-b}
    B_{l} = \sum_{j=j_{\min}}^{j_{\max}} \frac{\nu_j}{2} \operatorname{tr}  & \Big(  \mathbf{C}(f_j, \boldsymbol{\theta}_{\mathrm{bf}})^{-1}\frac{\partial \mathbf{C}(f_j)}{\partial \theta_l}\mathbf{C}(f_j, \boldsymbol{\theta}_{\mathrm{bf}})^{-1}  \nonumber \\
    & \operatorname{Bias}[\mathbf{\bar{P}}(f_j) \, | \, \boldsymbol{\theta}_{\mathrm{tr}}]\Big).
\end{align}
The spectrum bias is now a $3 \times 3$ matrix defined as the difference between the periodogram expectation and the spectrum matrix:
\begin{equation}
  \operatorname{Bias}[\mathbf{\bar{P}}(f_j) \, | \, \boldsymbol{\theta}_{\mathrm{tr}}] =  \operatorname{E}[\mathbf{\bar{P}}(f_j) \, | \, \boldsymbol{\theta}_{\mathrm{tr}}] - \mathbf{C}(f_j, \boldsymbol{\theta}_{\mathrm{tr}}).
\end{equation}
We apply this formalism to the case of \gls{sgwb} searches in \cref{sec:application-sgwb}.

\section{\label{sec:non-stat}Accounting for non-stationarity}

\subsection{Covariance and spectral estimate}

In this section, we extend the discussion to locally-stationary multivariate processes. Following~\cite{mallat_adaptive_1998}, one can describe a general Gaussian non-stationary process by its autocovariance function as
\begin{equation}
    \mathbf{R}(t, s) = \boldsymbol{\gamma}\left(\frac{t+s}{2}, t-s\right),
\end{equation}
where $\boldsymbol{\gamma}$ is a function of the middle time and the time difference of two time points $t$ and $s$. Note that for a multivariate process, $\mathbf{R}$ is a square matrix of dimension $p$, where $p$ is the number of channels.
We can then define an evolutionary spectrum as the Fourier transform of $\boldsymbol{\gamma}$:
\begin{align}
\label{eq:def-evolutionary-spectrum}
\mathbf{C}(u, f) & = \int_{-\infty}^{+\infty} \boldsymbol{\gamma}(u, v) e^{-2 i \pi f  v} d v.
\end{align}

Consider a time point, denoted by $\tau$. For a locally stationary process, the stationary assumption is only valid in the neighborhood of $\tau$, i.e., within an interval of size $l(\tau)$ around $\tau$. In this interval, we can make the approximation that the autocovariance function depends only on the difference between two time points. Formally, this statement can be expressed as follows:
\begin{align}
\label{eq:local-stationarity-assumption}
\forall t & \in [\tau - l(\tau) / 2,\, \tau + l(\tau)/2], \nonumber \\
    \mathbf{R}(t, s) & \approx \boldsymbol{\gamma}(\tau, t-s) \;\; \forall s.
\end{align}
The spectral consequence of this approximation is that for any point $(\tau, \xi)$ in the time-frequency plane, there is a rectangular box $I(\tau, \xi)$ around that point where the spectrum is approximately constant.
Hence, $\forall (t, f) \in I(\tau, \xi)$, we have $\mathbf{C}(t, f) \approx \mathbf{C}(\tau, \xi)$. This box is a rectangle of size $l(\tau) \times 1/d(\tau)$, where $d(\tau)$ is the decorrelation length (which verifies $d(\tau) < l(\tau)/2$).

Let us now consider the periodogram $P_L^{m}(f_k)$ defined in \cref{eq:time-segment-periodogram}, centered on some time point $t_m$, computed over a length $L$ equal to or smaller than the local stationary scale $l$. Similarly to Welch's periodogram expectation (see \cref{sec:time-binning}), it follows directly from \cref{eq:frequency-covariance-elements-multivariate} that
\begin{align}
\label{eq:short-term-periodogram-expectation}
\operatorname{E}\left[\mathbf{P}_{L\leq l}^{m}(f_k)\right] = \frac{1}{\tau_s \kappa_L} \int_{-\frac{f_s}{2}}^{+\frac{f_s}{2}} \left| \tilde{w}_L(f_k - f) \right|^2 \mathbf{C}(t_m, f) df,
\end{align}
which now depends on the time-varying spectrum matrix $\mathbf{C}(t_m, f)$. 

\subsection{Time-frequency approximate likelihood}

As for the stationary case, we can approximate the distribution of $\mathbf{P}_L^{m}(f_k)$ (or an averaged version of it) with a Wishart distribution:
\begin{align}
    \log f_{\mathbf{\bar{P}}} & = - \sum_{m, j} \frac{\nu_j}{2} \left(\log |\mathbf{C}(t_m, f_j)| \right. \nonumber \\
    & \left. + \operatorname{tr}\left(\mathbf{C}(t_m, f_j)^{-1} \mathbf{P}^{m}_{L}(f_j)\right)\right),
\end{align}
where we dropped the interval on which the time and frequency indices $m, j$ run for clarity. This means that the periodogram evaluated at a frequency bin $f_j$ computed over a time chunk centered on $t_m$ approximately follows a Wishart distribution with $\nu_j/2$ degrees of freedom, where the value of $\nu_j$ depends on the averaging that goes into the computation of the periodogram.

\subsection{Time-frequency bias estimate}

One can estimate the bias of any mismodeling of the spectrum $\mathbf{C}(t, f)$ of a locally stationary process by using the same formula as in~\cref{eq:parameter-bias}. The only difference is that one must replace $\mathbf{C}(f_j)$ by $\mathbf{C}(t_m, f_j)$ in \cref{eq:multivariate-fisher,eq:multivariate-b}.

A possible source of bias in time-frequency analysis is inappropriate time binning. This issue arises when computing periodograms over a length $L$ that is greater than the stationary timescale $l$. In such a situation, the time resolution is too coarse to account for short-scale variations in the spectrum. The resulting periodogram therefore reflects average behavior over the period on which it is computed. In that case, the periodogram expectation in~\cref{eq:short-term-periodogram-expectation} can be computed by replacing $\mathbf{C}(t_m, f)$ by an average over the considered period:
\begin{align}
\label{eq:long-term-periodogram-expectation}
\operatorname{E}\left[\mathbf{P}_{L > l}^{m}(f_k)\right] = \frac{1}{\tau_s \kappa_L} \int_{-\frac{f_s}{2}}^{+\frac{f_s}{2}} \left| \tilde{w}_L(f_k - f) \right|^2 \mathbf{C}_L(t_m, f) df,
\end{align}
where we set
\begin{equation}
    \mathbf{C}_{L}(t_m, f) = \frac{1}{L\tau_s}\int_{t_m-L\tau_s/2}^{t_m + L \tau_s/2} \mathbf{C}(t, f) dt.
\end{equation}
To estimate the bias due to a coarse gridding in time, one can use the difference between the periodogram expectation and the time-frequency covariance model evaluated at $t_m$:
\begin{equation}
\label{eq:spectrum-bias-coarse-time-gridding}
  \operatorname{Bias}[\mathbf{P}(t_m, f) \, | \, \boldsymbol{\theta}_{\mathrm{tr}}] =  \operatorname{E}\left[\mathbf{P}_{L > l}^{m}(f)\right] - \mathbf{C}(t_m, f).
\end{equation}
In a way similar to frequency-domain averaging, the bias depends on the second time derivatives of the spectrum. One can then directly plug \cref{eq:spectrum-bias-coarse-time-gridding} into \cref{eq:multivariate-b} to assess the bias on parameters controlling the covariance model. This is what we do in~\cref{sec:application-time-gridding-bias}.

\section{\label{sec:application-sgwb}Application to frequency binning in SGWB searches}

We apply the bias estimate presented in \cref{sec:inference} to assess the error due to binning in the frequency domain when searching for a \gls{sgwb} in \gls{lisa} data. 

\subsection{\label{sec:simulated-data}Simulated data}

We use the same \gls{lisa} noise configuration as in the Cosmo dataset \textsc{noise-1a}~\cite{bayle_2025_15698080} available on the \gls{ldc} website~\cite{ldc}. It includes a year of \gls{tdi} instrumental noise sampled at 0.5 Hz and generated with the package \textsc{LISA Instrument}~\cite{bayle_lisa_2023}. Spacecraft orbits are assumed to follow Keplerian orbits that minimize flexing to leading order in eccentricity, such that armlengths are approximately constant and equal.

We add an isotropic gravitational wave stochastic signal representing the radiation from cosmic strings. The parameter values are the same as in the dataset ``SGWB cosmic strings 1a'', or \textsc{cs-1a}, also downloadable from the \gls{ldc} website. Its all-sky all-polarization power spectral density is
\begin{equation}
\label{eq:sgwb-cosmic}
    S_h(f)= \Omega(f) \frac{3H_0^2}{4 \pi^2 f^3},
\end{equation}
where $\Omega(f)$ is the background energy density per logarithmic frequency parameterized as
\begin{equation}
    \Omega(f)=0.55 \times 10^{-11}\left(\frac{G\mu}{10^{-13}}\right)^{1 / 2}\left(\frac{f}{f_p}\right)^n,
\end{equation}
with $f_p = 3 \times 10^{-3}$~Hz, $n=0$ and $G\mu = 10^{-13}$.
In the \textsc{cs-1a} dataset, the \gls{sgwb} is simulated with \textsc{LISA GW Response}~\cite{bayle_2025_response}. Note that the simulation available in the \gls{ldc} website has a missing factor of $\sqrt{H_0}$ for the \gls{sgwb} amplitude, which we reapply to be consistent with \cref{eq:sgwb-cosmic}.

\subsection{Analysis covariance model}

We use a simple parametrized covariance model with three components: a low-frequency noise component, or \gls{tm} noise, a high-frequency noise component, or \gls{oms} noise, and a \gls{gw} background described by a power law, whose \gls{snr} peaks in the middle of the band. We assume that all component spectral shapes are known, but their respective amplitudes must be estimated.
The covariance model for the Michelson \gls{tdi} variables can be written as
\begin{equation}
\label{eq:covariance-model}
    \mathbf{C}_d(f) = \sum_{l} \alpha_l \mathbf{C}_l(f),
\end{equation}
where $\alpha_l$ are the \gls{psd} coefficients to be estimated, and the index runs on all components of the model $l=\{\mathrm{OMS},\mathrm{TM},\mathrm{GW}\}$. We build the noise covariance matrices $\mathbf{C}_l$ based on the model presented in~\cite{karnesis2026}, and the \gls{gw} covariances are built as in \cite{baghi_uncovering_2023}.

\begin{figure}
    \centering
    \includegraphics[width=1.0\columnwidth]{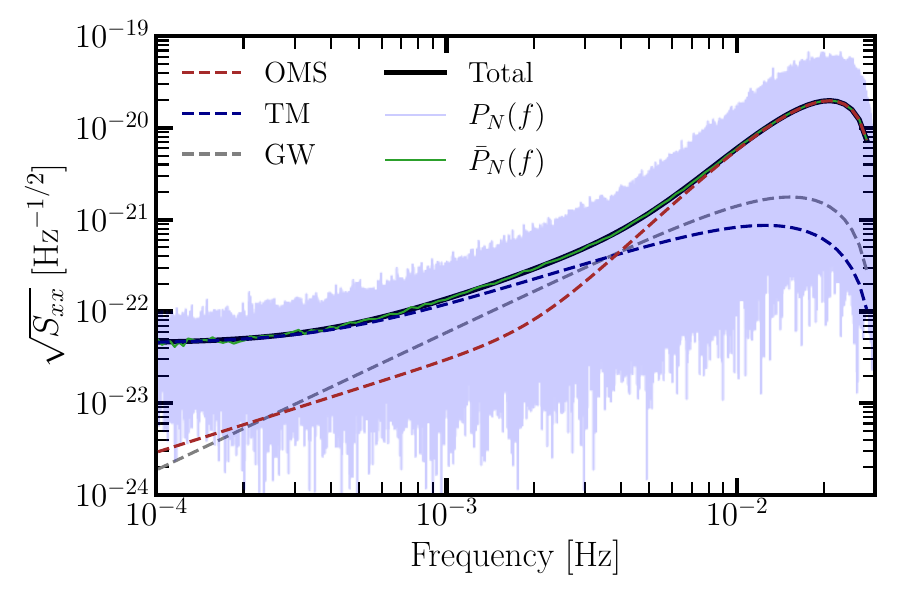}
    \caption{Components of the \gls{tdi} $X$ \gls{psd}: \gls{oms} noise (dashed red), \gls{tm} noise (dashed blue), \gls{sgwb} (dashed gray), and sum (solid black). The raw and smoothed periodograms of \gls{ldc} data are shown in light blue and green, respectively.}
    \label{fig:xx_spectra}
\end{figure}

As an illustration, we plot in~\cref{fig:xx_spectra} the three contributions of the first diagonal element of the covariance model (the \gls{psd} of $X$). The figure shows the \gls{oms} contribution, which mostly contributes to high frequencies, the \gls{tm} contribution that dominates low frequencies, and the \gls{sgwb} signal, which in this case reaches its highest \gls{snr} at mid-frequencies around 3 mHz. 

\subsection{\label{sec:application-averaging-bias}Forecast of the bias due to periodogram averaging}

After taking the \gls{dft} of the year-long time series, we select the bandwidth between $0.1$ and $28$ mHz, whose limits correspond respectively to the lower bound of the \gls{lisa} sensitive band and the first dip of the response with a safe margin. We then split the bandwidth into $J$ segments of constant size $N_j= N_{\mathrm{seg}}$ and form the binned periodogram $\bar{P}_N(f)$ following the averaging process outlined in \cref{sec:frequency-binning}. We define the averaging bandwidth as $N_{\mathrm{seg}} / T$, where $T$ is the observation time. Our goal is to choose an averaging bandwidth as large as possible for computational efficiency while keeping the induced bias below the statistical error.

\begin{figure}[htb]
    \centering
    \includegraphics[width=1.0\columnwidth]{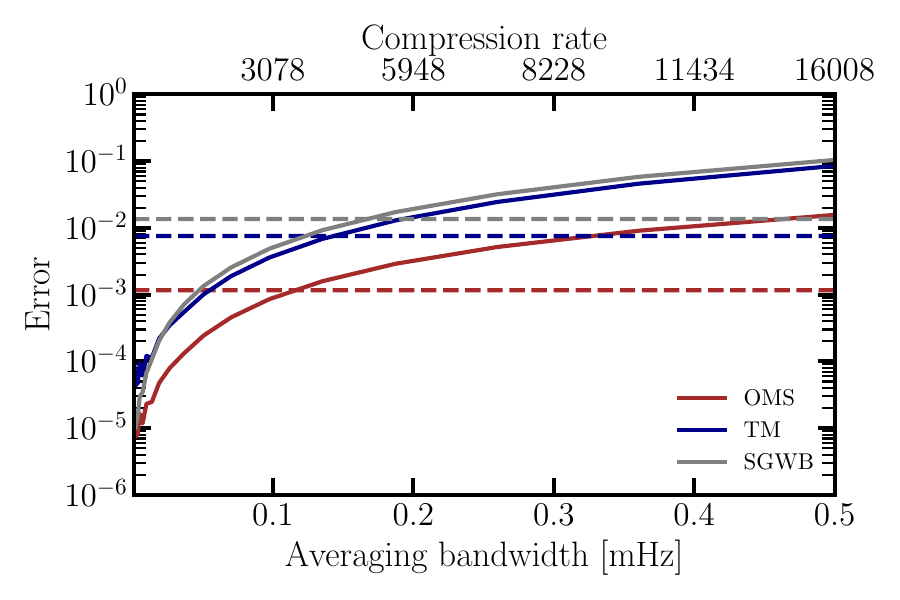}
    \caption{Bias and standard deviation of amplitude parameters as a function of the averaging bandwidth (lower axis) and compression rate (upper axis).}
    \label{fig:parameter-bias-variance}
\end{figure}

We test 20 averaging bandwidths logarithmically sampled from $10^{-6}$ Hz to $5 \times 10^{-4}$ Hz for which we evaluate the parameter standard deviations using~\cref{eq:multivariate-fisher} and their bias using~\cref{eq:parameter-bias}. They correspond to compression rates ranging from 31 to 16008 relative to the native frequency grid sampled every $1/T$. We report the results in \cref{fig:parameter-bias-variance}. 

Dashed lines represent standard deviations for \gls{oms} (red), \gls{tm} (blue), and GW (gray) amplitudes $\alpha_l$. As we analyze the same frequency range, they remain constant when we change the averaging bandwidth. Solid lines represent the amplitude bias, using the same color code. It increases with the averaging bandwidth. For each parameter, the closer the bias is to the standard deviation, the more significant it becomes compared to statistical errors. Based on this plot, one can state that using frequency segments smaller than $0.1$ mHz in width will result in negligible bias for all parameters. However, averaging over a width of $0.2$ mHz would induce a significant bias, at least for noise parameters. For an even more aggressive averaging, like $0.4$ mHz, the bias should be larger than the statistical errors for all parameters.

In \cref{fig:parameter-bias-variance}, we see that the bias increases monotonically with the averaging bandwidth. This is not necessarily a general behavior, as it could undergo local drops depending on the spectral shapes of the model components and how their second derivative varies with frequency.

We should note that the simple model in \cref{eq:covariance-model} that we adopt for illustration depends linearly on its parameters. In such a setting, we could have smoothed the model itself to avoid any bias and still benefit from the compression of the computational cost. Owing to the model linearity, we could have pre-computed the smoothed components $\bar{C}_l(f)$ and stored them in memory to avoid any additional cost in likelihood evaluations during sampling. However, \gls{sgwb} and noise models are generally more complex and include parameters controlling the spectral shape, for instance, when resorting to spline bases (e.g.,~\cite{baghi_uncovering_2023,karnesis2026}) or Gaussian processes~\cite{pozzoliWeaklyParametricApproach2024}, or when using stochastic background spectral templates (e.g.,~\cite{criswell_templated_2024,boileau_spectral_2021}). In that case, pre-computing is not possible. The only way to achieve compression is to optimize the smoothing bandwidth through the method we propose, so that the bias remains under control.

\subsection{\label{sec:posterior-sampling}Bias verification with posterior sampling}

We test the bias forecast of \cref{fig:parameter-bias-variance} by running a Bayesian inference of the three parameters $\alpha_l$ with uniform priors in the range $[-2, 2]$. We use the Wishart likelihood in~\cref{eq:wishart_loglike}. Note that the choice of priors is usually critical in \gls{sgwb} searches or in Bayesian spectrum estimation in general, as discussed in Ref.~\cite{armanoIndepthAnalysisLISA2024}. We set this issue aside in the present work to concentrate on the influence of periodogram smoothing, and we investigate this issue in detail in another study~\cite{karnesis2026}. Besides, in our inference, the frequency dependence of the different components is fixed, which attenuates the impact of priors.

We perform two inferences: in the first one, we use ``noiseless'' data: the data periodogram matrix $\mathbf{\bar{P}}(f)$ in the likelihood is taken equal to its statistical expectation $\operatorname{E}[\mathbf{\bar{P}}(f)]$. This provides a reference that is independent of statistical fluctuations. In the second inference, we directly use the periodogram matrix computed from noise and signal realizations obtained by combining the \textsc{noise-1a} and \textsc{cs-1a} data of the \glspl{ldc}. The injection corresponds to parameter values $\alpha_l = 1$ for all components $l$. We show the injected raw data periodogram in \cref{fig:xx_spectra}, and the smoothed periodogram which is used in the likelihood in green. 
We perform the inference for three choices of averaging bandwidth: $0.05$ mHz, $0.2$ mHz, and $0.4$ mHz. These are referred to as high, medium, and low resolution, respectively.

\begin{figure}
    \centering
    \includegraphics[width=1.0\columnwidth]{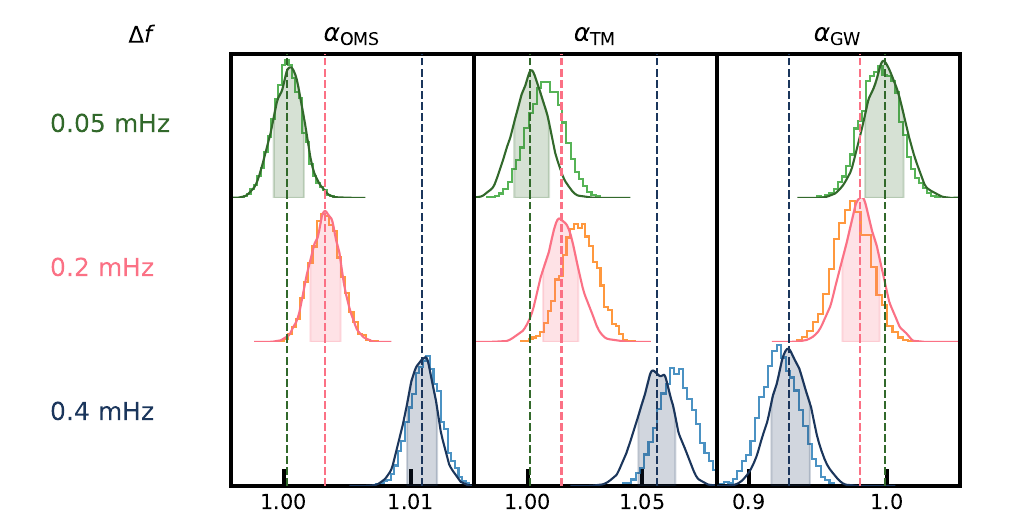}
    \caption{From top to bottom, the rows correspond to marginalized parameter posteriors for averaging bandwidths of $0.05$ (green), $0.2$ (red), and $0.4$ mHz (blue). From left to right, the three columns correspond to coefficients for the \gls{oms}, \gls{tm}, and GW components in the covariance. Vertical dashed lines are biased values predicted by \cref{eq:parameter-bias}, to be compared with the injected value which is $\alpha_l=1$ for all parameters. The smoothed \glspl{pdf} represent the zero-noise posteriors, while the histograms show the distributions obtained from \gls{ldc} data simulations. This plot was realized with \textsc{ChainConsumer}~\cite{Hinton2016}.}
    \label{fig:biased-posteriors}
\end{figure}

We use the \textsc{eryn} ensemble sampler~\cite{karnesis2023,michael_katz_2023_7705496,Foreman-Mackey2013} with 12 walkers to explore the posterior distribution. After a burn-in phase of 2000 steps, we collect 500 posterior samples by recording every fifth step. We plot the marginalized posteriors in \cref{fig:biased-posteriors} for the three cases of averaging bandwidth. Each column corresponds to one parameter, while each row (and color) corresponds to an averaging bandwidth. The vertical dashed lines indicate the values $\alpha + \Delta \alpha$ where $\Delta \alpha$ is the bias predicted by \cref{eq:parameter-bias}. The smooth, solid line distributions represent ``noiseless'' posteriors, whereas step histograms represent posteriors obtained with one data realization (\gls{ldc} data). All lines are aligned with the smooth posterior modes, confirming the reliability of the bias forecast at first order. The posteriors obtained with the data realization are consistent with the ``noiseless'' ones within statistical errors. This means that the analytical modeling of the covariance agrees with the simulated data, even though the latter was generated independently using time-domain simulators.

The high-resolution case (green) exhibits a negligible bias since both vertical lines and posteriors are almost aligned with the injected values $\alpha_l =1$. Departures from one become visible for all parameters when using the medium resolution (red), but a bias would be hardly distinguishable from statistical errors, especially for the \gls{gw} component. Using the low resolution (dark blue) induces statistically significant biases, in accordance with the forecast.

\section{\label{sec:application-time-gridding-bias}Application to time binning in SGWB searches}

\subsection{Data simulation} 

In this section, we consider the same instrumental setup as in \cref{sec:simulated-data} except that we use numerically optimized spacecraft orbits featuring larger time variations. The simulation of the spacecraft positions can be retrieved from a repository made available by ESA~\footnote{\url{https://github.com/esa/lisa-orbit-files}}. The corresponding orbit file is produced by the software \textsc{LISA Orbits}~\cite{bayle_2025_orbits}.

\begin{figure}
    \centering
    \includegraphics[width=1.0\columnwidth]{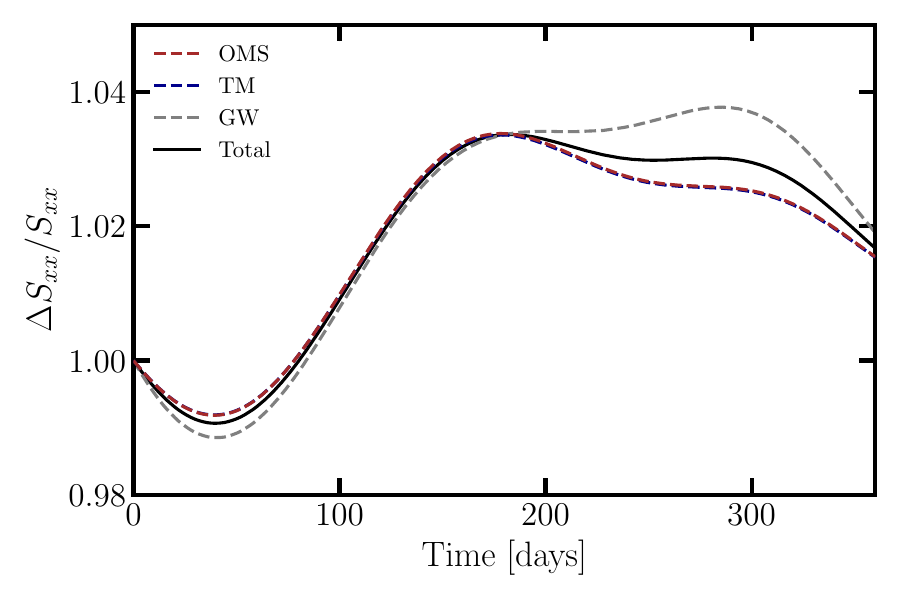}
    \caption{Variation of \gls{tdi} $X$ \gls{psd} relative to its initial value as a function of time, evaluated at frequency $f = 3$ mHz. We break it into three components: \gls{oms} noise (dashed red), \gls{tm} noise (dashed blue), \gls{sgwb} (dashed gray). The relative variation of the total \gls{psd} is shown by the solid black line.}
    \label{fig:s_xx_variation}
\end{figure}

We can compute the instrumental noise and \gls{sgwb} covariance matrices from the orbits. Even though the stochastic sources (\gls{sgwb}, \gls{oms}, and \gls{tm} noises) are stationary, the dependence of the \gls{tdi} transfer functions on the light travel times between spacecraft induces a slow non-stationarity in the measured process. We illustrate this by plotting the relative variation of \gls{tdi} $X$ \gls{psd} divided by its value at the initial time in \cref{fig:s_xx_variation} at 3 mHz. We observe slow variations of about 5\% over one year. Note that the relative variation is different depending on the considered frequency.

\subsection{Forecast of the bias due to coarse time gridding}

\begin{figure}
    \centering
    \includegraphics[width=\columnwidth]{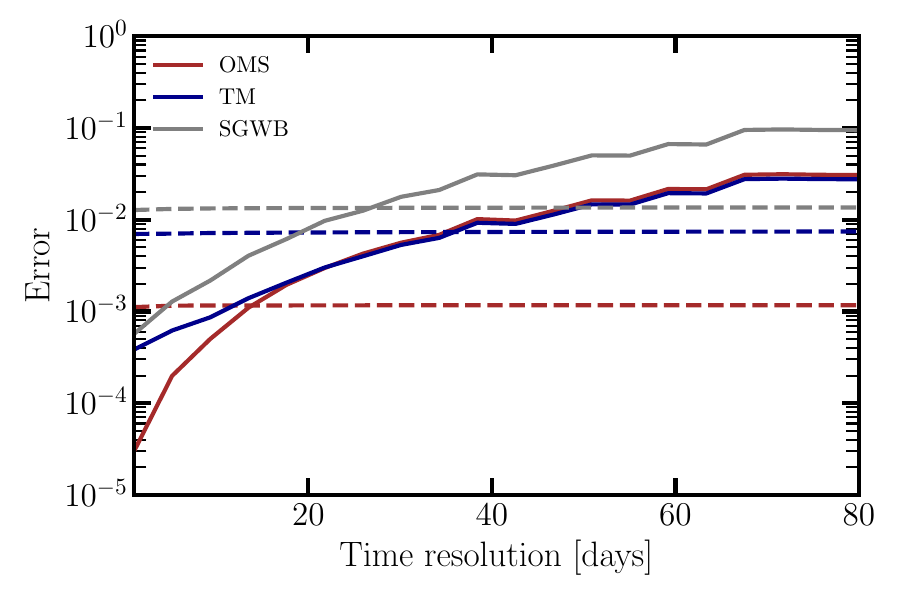}
    \caption{Bias on spectrum component amplitudes as a function of the resolution of the time chunking for \gls{oms} noise (red), \gls{tm} noise (blue), and \gls{sgwb} (gray). The solid lines show the parameter biases, while the dashed lines show their standard deviations.}
    \label{fig:bias-vs-time-resolution}
\end{figure}

We investigate the effect of time resolution on the accuracy of $\alpha_l$ parameter inference. We build several time grids $t_m$ starting from day-long to 80-day-long time chunks. For each of these chunks, we compute the covariance model $\mathbf{C}(t_m, f)$ and the periodogram expectation $\operatorname{E}\left[\mathbf{P}_{L}^{m}(f)\right]$ using \cref{eq:long-term-periodogram-expectation}. We compute the parameter bias using \cref{eq:parameter-bias} and report the results in \cref{fig:bias-vs-time-resolution}.

The bias remains below or comparable to the statistical error (a few $\sigma$) for chunks that are smaller than 20 days. Above this threshold, the bias starts to be significant relative to the standard deviation, especially for the \gls{oms} noise amplitude. For resolutions of 80 days, the bias on the \gls{sgwb} amplitude reaches nearly 10 times the standard deviation in absolute value. This indicates that accurate inference with \gls{lisa} data must account for the time variations of the spectrum of noise and signal components, even if the underlying process is originally stationary (and even if the signal source is isotropically distributed over the sky).

\begin{figure}
    \centering
    \includegraphics[width=\columnwidth]{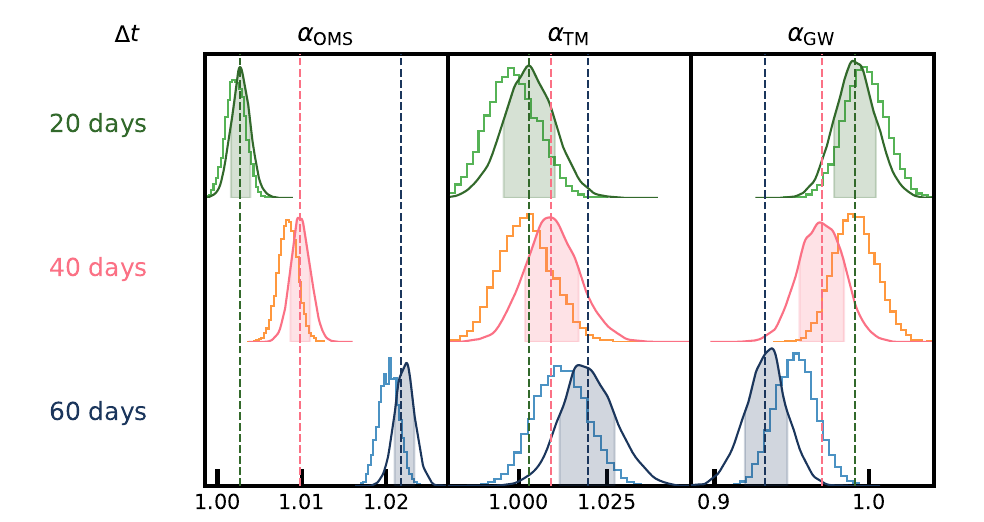}
    \caption{\label{fig:bias-vs-time-resolution-posteriors}From top to bottom, the rows correspond to marginalized parameter posteriors for time resolutions of 20 (green), 40 (red), and 60 days (blue). From left to right, the three columns correspond to coefficients for the \gls{oms}, \gls{tm}, and GW components in the covariance. Vertical dashed lines are biased values predicted by~\cref{eq:parameter-bias}, to be compared with the injected value which is $\alpha_l=1$ for all parameters. The smoothed \glspl{pdf} represent the zero-noise posteriors, while the histograms show the distributions obtained from the \gls{ldc} data simulations.}
\end{figure}

We verify this behavior by sampling the posterior using the same uninformative uniform distributions as in \cref{sec:posterior-sampling} for a selection of 3 time resolutions: 20 days (high resolution), 40 days (medium), and 60 days (low). We plot the marginalized posteriors of amplitude parameters in \cref{fig:bias-vs-time-resolution-posteriors}. Again, the smooth \glspl{pdf} represent ``noiseless'' inferences, while the bar histograms are the results obtained with the \gls{ldc} data realization. 

We observe an agreement between the parameter posterior means and the value that the bias estimation predicts, demonstrating the reliability of the approximation for this model. Additionally, the posterior distribution we obtain from the \gls{ldc} data realization (bar histograms) is also statistically consistent with the posterior obtained from the noiseless likelihood. In the case of high (green) and medium (red) resolutions, the bias is hardly distinguishable from the statistical error, except for the \gls{oms} noise amplitude (left panel). The low-resolution analysis (blue) exhibits a clear bias for all parameters, in agreement with the analytical forecast. 
This is confirmed by the analysis performed on the independently generated \gls{ldc} datasets (bar histograms in \cref{fig:bias-vs-time-resolution-posteriors}), where the posterior distributions are in statistical agreement with the ``noiseless'' posteriors.

Therefore, an analysis that would assume stationarity, or even a time-frequency analysis with an inappropriate time binning, would induce a statistically significant error on isotropic \gls{sgwb} parameter inference only because of the time variations of the instrument response.



\section{\label{sec:discussion}Discussion}

When characterizing the spectrum of stationary or locally stationary time series, it is often useful to define sufficient statistics, i.e., transform the data into a collection of reduced quantities (spectral estimates) that contain the necessary information while being more compact than the original data. 
We reviewed how correlations between adjacent bins may affect the distribution of spectral estimates. When averaging several periodograms, the statistics of the resulting estimate depend on correlations among periodogram ordinates. 
The effect of correlations can be approximately accounted for by assuming that the spectrum is a random variable proportional to a chi-squared distribution in the univariate case, or a Wishart distribution in the multivariate case. The number of degrees of freedom of this distribution is set by the number of averaged segments (either in time or frequency) and by the level of correlations between bins. The correlation relates to the shape of the taper window and its potential overlap between bins.

We argued that frequency-domain smoothing typically provides greater data compression than time-segment averaging. It also has the advantage of being less susceptible to leakage because the time window size is larger. However, frequency smoothing introduces a bias related to spectral variations across the averaged frequency bandwidth. To mitigate this effect, we showed how to predict the bias in order to achieve the optimal balance between data compression and accuracy. For spectra depending on unknown parameters, we derived a simple estimate of the parameter bias due to frequency binning based on a Taylor expansion up to first order in spectrum derivatives. We gave an example of an application with a simulation of \gls{lisa} data featuring a mixture of instrumental noises and a \gls{sgwb}. We were able to determine for which binning the bias can be neglected compared to the statistical error.

Furthermore, we demonstrated that the same bias estimation framework can be applied to handle spectral non-stationarities. In time-frequency data analyses based on the short-time Fourier transform, a time resolution must be chosen that is fine enough to describe the characteristic timescale of the underlying stochastic process. This scale determines the period over which the process can be considered approximately stationary. Coarser time resolutions typically reduce computational complexity and increase the sensitivity to lower frequencies. However, if the resolution is low compared to the stationary length, this directly results in a bias when estimating the spectrum. 
Studies addressing non-stationarity in LISA data span a broad range of time resolutions. For instance, \citet{mentastiProbingGalacticExtragalactic2024} adopt a short, one-day resolution to remain sensitive to anisotropies in the Galactic and potentially extragalactic backgrounds. This selection reflects a trade-off between accessing sufficiently low frequencies and tracking how rapidly the observed sky evolves. Likewise, \citet{buscicchioTestLISAForeground2025} use one-day segments to carry out Gaussianity and stationarity tests on LISA data, and emphasize that the outcomes vary with the chosen time-chunk duration. In contrast, \citet{adams_detecting_2014} and \citet{hindmarshRecoveringPhaseTransition2025} aim to recover a primordial SGWB in the presence of the Galactic foreground and employ a coarser resolution of roughly one week, which performs well in practice though without a quantitative justification. \citet{rosatiPrototypeStochasticGravitational2025} instead split the data into longer, one-month intervals after empirically evaluating several segment lengths. Here, we propose a systematic procedure to identify, for a specified model, the segmentation that best balances model complexity and inference accuracy when non-stationary processes are present.

We showed that, assuming realistic constellation orbits and a model with free amplitudes for both noise and signal, the time resolution needs to be shorter than 20 days to prevent any detectable bias arising from annual modulations of \gls{tdi} transfer functions across one year of data. This finding, however, pertains only to arm flexing: modulation of the Galactic foreground and potential intrinsic noise non-stationarities will likely require even stricter segmentation.

Our analysis demonstrates that performing accurate \gls{sgwb} searches with detector data requires fine control over correlation effects in time and frequency. We propose a Fisher-information-based tool that provides a method for selecting adapted resolutions based on the expected statistical precision. Beyond binning effects, our analysis can be used to quantify parameter biases arising from any discrepancy between the data and the model.

In this work, we considered the effect of correlations on the statistics of averaged spectral estimates. Although we expect it to be marginal, we ignored the correlations among neighboring averaged periodograms in time or frequency, which allowed us to write down a likelihood that is the product of single-frequency or time-frequency bin likelihoods. In the future, we plan to study the significance of these correlations in the context of \gls{sgwb} searches, and explore the advantage of using different time-frequency representations like wavelets.

\section*{Acknowledgments}
\label{sec:Acknowledgments}

We acknowledge very insightful discussions with R. Buscicchio, O. Burke and A. Vajpeyi. We also thank F. Pozzoli, M. Muratore, N. Dam Quang, O. Hartwig, M. Pieroni, A. Santini, R. Meyer, S. Babak, C. Caprini, H. Inchauspé, A. Petiteau, M. Besançon, and G. Nardini and all the members of the \gls{lisa} \gls{sgwb} code comparison team. We are grateful to C. Cavet who keeps the \gls{ldc} website up and running. NK was supported by the Hellenic Foundation for Research and Innovation (H.F.R.I.) under the 4th Call for HFRI Research Projects to support Post-doctoral Researchers (Project Number: 28418). This work was performed using HPC resources from CNES Computing Center.

\section*{Software and data}
\label{sec:software}

We have used the {\tt backgrounds}~\footnote{\url{https://gitlab.in2p3.fr/qbaghi/backgrounds}} software to perform the inferences and model signal and instrumental responses. The notebooks to reproduce the results are also available~\footnote{ \url{https://gitlab.in2p3.fr/qbaghi/backgrounds-papers}}. We used the {\texttt{Eryn}}~\footnote{\url{https://github.com/mikekatz04/Eryn}} sampler to probe the posterior distributions using Markov chain Monte Carlo. The datasets used in this work can be found on the \gls{ldc} website~\cite{ldc,bayle_2025_15698080}.

\appendix

\section{\label{sec:time-binning}Effect of binning in the time domain}

In this section, we investigate the effect of averaging periodograms of several time segments, also called Welch's method~\cite{welch_use_1967}, on the likelihood function.

\subsection{Approximate distribution of spectral estimates}

Welch's method consists of partitioning the time series into $M$ segments of length $L$ that are potentially overlapping, and averaging their periodograms. Similarly to Eq.~\eqref{eq:wdft}, we can define the (tapered) \gls{dft} of the $m^{\text{th}}$ segment $\forall m \in [0;\; M-1]$:
\begin{equation}
\label{eq:segment_dft}
    \tilde{x}^{m}(f_k) = \tau_s \sum_{n=0}^{L-1} w_n x_{(n+ms)} e^{-2 i \pi \frac{nk}{N}},
\end{equation}
where $s \in ]0;\; L]$ is the shift in samples between two consecutive segments. By defining $p = (L-s)/L$ as the percentage of overlap between two consecutive segments, we have the following relationship between $M$, $L$ and $p$~\cite{astfalck_debiasing_2024}:
\begin{equation}
    ML(1-p) + Lp = N.
\end{equation}
Welch's \gls{psd} estimator computes the average of the segment periodograms as
\begin{equation}
    \bar{P}_{L}(f_k) = \frac{1}{M} \sum_{m=0}^{M-1} P^{m}_{L}(f_k),
\end{equation}
where $P^{m}_{L}(f_k)$ is the periodogram of segment $m$ evaluated at frequency $f_k$:
\begin{equation}
\label{eq:time-segment-periodogram}
    P^{m}_{L}(f_k) \equiv \frac{2}{\tau_s \kappa_L}| \tilde{x}^{m}(f_k) |^2.
\end{equation}
The expectation of Welch's periodogram is equal to the diagonal element of the frequency-domain covariance of a segment of length $L$, as introduced in Eq.~\eqref{eq:frequency-covariance-elements}:
\begin{align}
\label{eq:welch-expectation}
\operatorname{E}\left[\bar{P}_{L}(f_k)\right] & = \frac{2}{\tau_s \kappa_L}\tilde{\Sigma}_L(k, k) \nonumber \\
& = \frac{1}{\tau_s \kappa_L} \int_{-\frac{f_s}{2}}^{+\frac{f_s}{2}} \left| \tilde{w}_L(f_k - f) \right|^2 S_{xx}(f) df.
\end{align}
For convenience, we will denote this expectation by 
\begin{equation}
    \bar{S}_L(f_k) \equiv \operatorname{E}[\bar{P}_{L}(f_k)]
\end{equation}
in the following. When the windowing is adapted, and $L$ is sufficiently large, it can be approximated by the true \gls{psd} $\bar{S}_L(f_k) \approx S(f_k)$.

To compute the variance, one has to account for cross-covariances between segment periodograms. 
We have~\cite{astfalck_debiasing_2024}:
\begin{align}
\label{eq:welch-variance-approx-1}
    \operatorname{Var}\left[\bar{P}_L(f_k)\right] & \approx \frac{\operatorname{Var}\left[P_L(f_k)\right]}{M} \Bigg( 1 + \frac{2}{M} \sum_{m < m'}^{M-1} c_{mm'}(f_k) \Bigg)
\end{align}
where $c_{mm'}$ is the correlation coefficient of two segment periodograms:
\begin{equation}
\label{eq:segment-correlation}
    c_{mm'}(f_k) \equiv \frac{\operatorname{Cov}[P^{m}_{L}(f_k), P^{m'}_{L}(f_k)]}{\sqrt{\operatorname{Var}[P_L^{m}(f_k)] \operatorname{Var}[P_L^{m'}(f_k)]}}.
\end{equation}
Note that $\operatorname{Var}[P_L^m(f_k)] = \operatorname{Var}[P_L^{m'}(f_k)] \approx \bar{S}_L(f_k)^2$ for all $m$ (the segment periodograms have all the same variance, approximately equal to their squared expectation). 
From \cref{eq:welch-variance-approx-1} we see that the variance of the averaged segment periodograms is not simply the variance of a single periodogram divided by $M$. 
Due to correlations among segment periodograms $P^{m}_L(f_k)$, there is a penalizing term that increases it. Consequently, in general, Welch's estimate does not simply follow a chi-squared distribution with $2 M$ \glspl{dof} as it would if segments were uncorrelated. Rather, the statistical distribution of $\bar{P}_L(f_k)$ can be approximated to be proportional to a chi-squared distribution with an effective number of \glspl{dof} $\nu < 2M$ that accounts for segment-to-segment correlations~\cite{percival_spectral_2020}, as we did in Sec.~\ref{sec:frequency-binning}:
\begin{equation}
\label{eq:pdf-propto-chi2-welch}
    \bar{P}_L(f_k) \sim a \chi^2_{\nu}.
\end{equation}
In order for this distribution to have the same mean and variance, we require $\nu$ to be~\cite{noauthor_psd_1991}
\begin{equation}
\label{eq:effective-ndof-1}
    \nu \approx \frac{2M}{1  + \frac{2}{M} \sum_{m < m'}^{M-1} c_{mm'}(f_k) }.
\end{equation}
and $a$ to be
\begin{equation}
    a = \frac{\bar{S}_L(f_k)}{\nu}.
\end{equation}

As shown in~\cref{sec:time-correlations}, the correlation coefficients mainly depend on the window, and can be approximated as
\begin{equation}
\label{eq:correlation-coefficient-welch}
    c_{mm'}(f_k) \approx \frac{1}{\kappa_L^2}\left| \sum_{n=0}^{L-1} w_n w_{n + s|(m'-m)|}\right|^2,
\end{equation}
which greatly simplifies the computation of the sum in \cref{eq:effective-ndof-1}.

We verify that if we ignore correlations ($c_{mm'} = 0$ for $m \neq m'$), then the effective number of \glspl{dof} reduces to $2M$. Note that assuming the averaged periodogram is proportional to a chi-squared random variable is equivalent to assuming that it follows a gamma distribution. This modeling approach is adopted and justified in~\cite{Nobili2026FastPreMerger} in the context of massive black hole detection with \gls{lisa}.

\subsection{\label{sec:example-time-averaging}Example using simulated samples}

As for frequency averaging, we illustrate the effect of correlations in time binning with a numerical experiment. We consider the same simulation set as in \cref{sec:example-frequency-averaging}, with the same number of realizations.

\begin{figure}
\begin{tabular}{c}
     \includegraphics[width=\columnwidth]{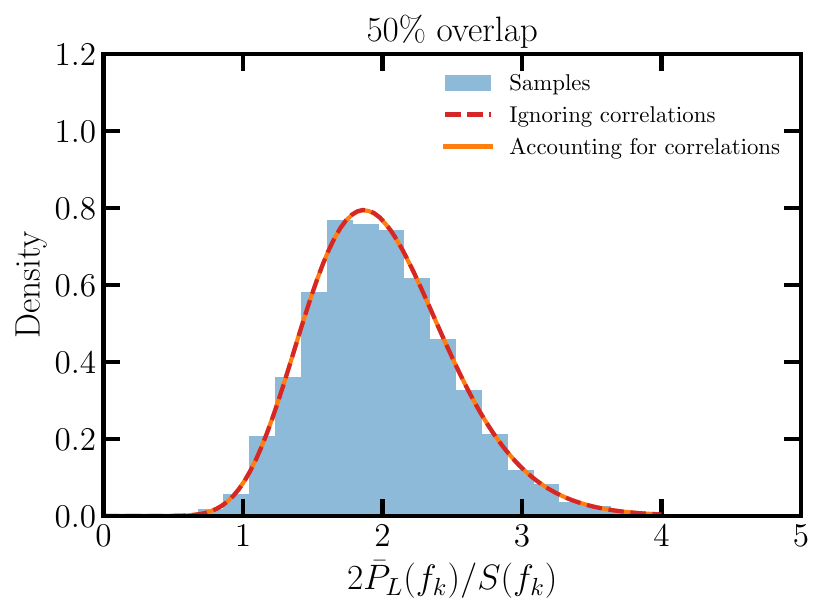}\\
     \includegraphics[width=\columnwidth]{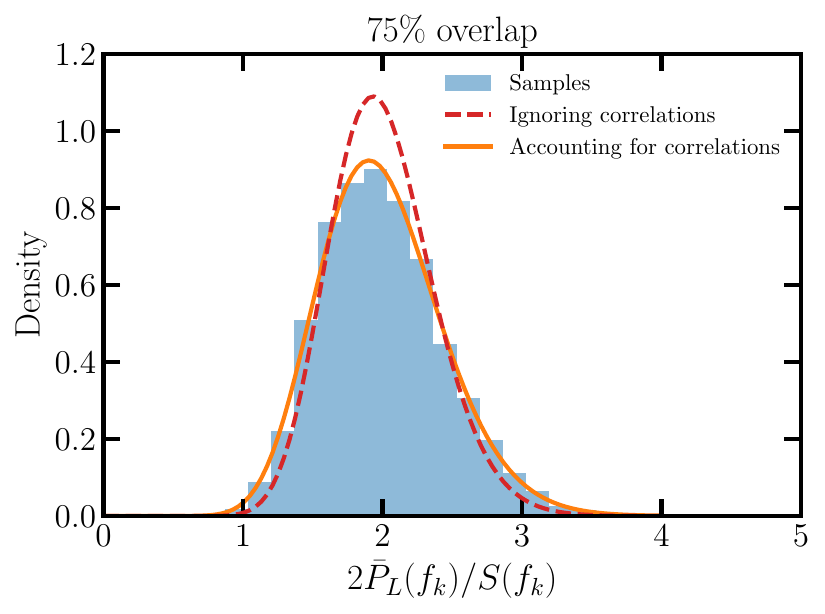}
\end{tabular}
 \caption{\label{fig:welch-periodogram}Blue histograms: Welch's periodogram distribution using $10^4$ realizations of the same noise PSD at $f = 1~\mathrm{mHz}$ with segment overlaps of $p=50\%$ (upper panel) and $p=75\%$ (lower panel). Lines: distributions modeled with~\cref{eq:pdf-propto-chi2-welch} with $2M$ \glspl{dof} (dashed red) and $\nu$ \glspl{dof} (solid orange). Segment correlations start contributing with overlaps larger than 50~\%, and ignoring correlations leads to underestimating the variance.}
\end{figure}

For all realizations, we compute Welch's periodogram with a segment size of $L = 1024$ and a Blackman-Harris window using two percentages of overlap $p=0.5$ and $p=0.75$. We plot the distribution of the periodogram evaluated at a single frequency $f = 1$ mHz in \cref{fig:welch-periodogram} for the two choices of overlaps.

In the case of 50~\% overlap (top panel), which corresponds to the separation between two maxima of the window, the distribution (orange) is indistinguishable from a chi-squared distribution with $2M$ degrees of freedom (dashed red), as we would have for perfectly uncorrelated periodograms. Increasing the overlap to 75\% (lower panel) breaks this approximation, and the distribution that ignores correlations significantly deviates from the one with an adjusted number of \glspl{dof}.


\section{\label{sec:lisa-psd}Baseline instrumental noise PSD}

The instrumental noise \gls{psd} used in this work assumes two independent noise sources coming from test-mass (\gls{tm}) free-fall perturbations affecting low frequencies and optical metrology system (\gls{oms}) noise affecting high frequencies. Their spectral shapes follow the technical note LISA-LCST-SGS-TN-001~\cite{babak2021lisasensitivitysnrcalculations} derived from the LISA Science Requirement (SciRD), but their amplitudes differ from the SciRD, to match the simulations in the \gls{ldc} Cosmo dataset. For the test-mass noise we have
\begin{align}
	S_{\mathrm{TM}}(f) = a_{\mathrm{TM}}^2 \left[1 + \left(\frac{f_{1}}{f}\right)^2 \right] \left[1 + \left(\frac{f}{f_2}\right)^4\right],
\end{align}
where $a_{\mathrm{TM}} = 2.4 \times 10^{-15}\,  \mathrm{m\,s^{-2}}$, $f_{1} = 4 \times 10^{-4}$ Hz and $f_2 = 8$ mHz.
For the \gls{oms} readout noise, we have
\begin{align}
	S_{\mathrm{OMS}}(f)= a_{\mathrm{OMS}}^2 \left[1 + \left(\frac{f_3}{f}\right)^4\right],
\end{align}
where $a_{\mathrm{OMS}} = 7.9 \times 10^{-12} \, \mathrm{m\,Hz^{-1/2}}$ and $f_3 = 2$~mHz.

\section{\label{sec:time-correlations}Correlations of periodograms of different time segments}

In this section, we compute the correlation coefficient between periodograms computed on two different segments $m$ and $m'$, but at a single frequency $f_k$, as defined in \cref{eq:segment-correlation}. We reproduce its definition here:
\begin{equation}
    c_{mm'}(f_k) \equiv \frac{\operatorname{Cov}\left[P^{m}_{L}(f_k), P^{m'}_{L}(f_k)\right]}{\operatorname{Var}\left[P_L(f_k)\right]}.
\end{equation}

We start with the denominator. For large enough $L$, the variance of a segment periodogram is approximately equal to its squared expectation (see, e.g., \cite{priestley_spectral_1982} or \cite{krogstad_covariance_1982}):
\begin{equation}
\label{eq:segment-variance}
    \operatorname{Var}\left[P_L(f_k)\right] \approx \operatorname{E}\left[P_{L}(f_k)\right]^2 \approx S_{xx}(f_k)^2.
\end{equation}
Now we consider the numerator of~\cref{eq:segment-correlation}. By using the definition of Welch's periodogram in \cref{eq:time-segment-periodogram}, and using Isserlis' theorem~\cite{Isserlis1918}, one can show that for a zero-mean Gaussian process, the covariance of two spectral estimates evaluated at the same frequencies but computed from two different segments $m$ and $m'$ is
\begin{align}
\label{eq:segment-periodogram-covariance}
    \operatorname{Cov}\left[P_L^m(f_k), P_L^{m^{\prime}}(f_k)\right]  = \left(\frac{2}{\tau_s \kappa_L}\right)^2 \Big( & \left| \operatorname{E} [\tilde{x}^{m}_k \tilde{x}^{m^{\prime}\ast}_k ] \right|^2 \nonumber \\
    & + \left| \operatorname{E} [\tilde{x}^{m}_k \tilde{x}^{m^{\prime}}_k ]\right|^2 \Big) .
\end{align}
This result is recalled in~\cite{percival_spectral_2020}, p. 442. The expectation in the first term in \cref{eq:segment-periodogram-covariance} is the covariance of two segment \glspl{dft} and can be written as:
\begin{align}
\label{eq:first-term}
    \operatorname{E} [\tilde{x}^{m}_k \tilde{x}^{m^{\prime}\ast}_k ]  = \frac{1}{2} \int_{-\frac{f_s}{2}}^{+\frac{f_s}{2}} & \left| \tilde{w}_L(f_k - f) \right|^2 \nonumber \\
    &  \times S_{xx}(f) e^{2 i \pi f s (m^{\prime} - m) \tau_s} df.
\end{align}
Similarly, the expectation in the second term in \cref{eq:segment-periodogram-covariance} writes
\begin{align}
\label{eq:second-term}
    \operatorname{E} [\tilde{x}^{m}_k \tilde{x}^{m^{\prime}}_k]  = \frac{1}{2} \int_{-\frac{f_s}{2}}^{+\frac{f_s}{2}} & \tilde{w}_L(f_k - f) \tilde{w}_L(f_k + f) \nonumber \\
   & \times S_{xx}(f) e^{2 i \pi f s (m^{\prime} - m) \tau_s} df.
\end{align}
Note the difference with \cref{eq:frequency-covariance-elements}: now the expression includes the time shift between the two segments.
Now we make two approximations. First, far from the Nyquist and zero frequencies, the second term (\cref{eq:second-term}) can be neglected compared to the first term (\cref{eq:first-term})~\cite{thomson_spectrum_1977}, due to the fact that the window \gls{dft} is strongly peaked around zero. Second, if we assume that the \gls{psd} is approximately constant in the bandwidth within the maximum of the window $\tilde{w}_{N}(f)$, then we can get it out of the integrals. This gives
\begin{align}
    \operatorname{E} [\tilde{x}^{m}_k \tilde{x}^{m^{\prime}\ast}_k ] & \approx \frac{1}{2} S_{xx}(f_k) \int_{-\frac{f_s}{2}}^{+\frac{f_s}{2}} | \tilde{w}_L(f_k - f) |^2.
\end{align}
Replacing the window \gls{dft} $\tilde{w}_L$ by its definition leads to
\begin{equation}
\label{eq:segment-periodogram-covariance-approx}
    \operatorname{Cov}[P_L^m(f_k), P_L^{m^{\prime}}(f_k)] \approx
    \frac{S_{xx}(f_k)^2}{\kappa_L^2}\left| \sum_{n=0}^{L-1} w_n w_{n + s|(m'-m)|}\right|^2 
\end{equation}
where in the equation above we assume that $w_n = 0$ for $n \geq L$. 
Using \cref{eq:segment-periodogram-covariance-approx} and \cref{eq:segment-variance}, \cref{eq:segment-correlation} becomes
\begin{equation}
    c_{mm'}(f_k) \approx \frac{1}{\kappa_L^2}\left| \sum_{n=0}^{L-1} w_n w_{n + s|(m'-m)|}\right|^2 
\end{equation}
Now we can compute the sum in \cref{eq:effective-ndof-1}:
\begin{align*}
\sum_{m<m^{\prime}}^{M-1} c_{mm'}(f_k) & = \frac{1}{\kappa_L^2} \sum_{m<m^{\prime}}^{M-1}\left| \sum_{n=0}^{L-1} w_n w_{n + s|(m'-m)|}\right|^2  \\
&= \sum_{p=1}^{M-1}\left(M - p\right) \frac{1}{\kappa_L^2}  \left| \sum_{n=0}^{L-1} w_n w_{n + ps}\right|^2 \\
& = \sum_{p=1}^{M-1}\left(M - p\right)\rho(p, s),
\end{align*}
where we converted the double sum on $m$ and $m'$ into a single sum by setting $p=m'-m$, and we defined the normalized window correlation function as in Ref.~\cite{noauthor_psd_1991}:
\begin{equation}
\label{eq:defrho}
    \rho(p, s) \equiv 
    \begin{cases}
            \frac{1}{\kappa_L^2} \left| \sum_{n=0}^{L-1-ps} w_n w_{n + ps} \right|^2 & 1 \leq p \leq \lfloor \frac{M}{s} \rfloor \\
           0 & \text{otherwise.}
    \end{cases}
\end{equation}
Plugging this result into \cref{eq:effective-ndof-1} leads to \begin{equation}
\label{eq:effective-ndof-2}
    \nu \approx \frac{2M}{1  + 2 \sum_{p=1}^{M-1} \left(1 - \frac{p}{M} \right) \rho(p, s) }.
\end{equation}

\section{\label{sec:frequency-correlations}Correlations of periodograms at different frequencies}

In this section, we derive the expression of the correlation coefficient between the periodograms evaluated at two different frequencies, defined as
\begin{equation*}
		\sigma_{kk'}  \equiv 
		\frac{\operatorname{Cov}[P_N(f_k), P_N(f_{k'})]}{\sqrt{\operatorname{Var}[P_{N}(f_k)]\operatorname{Var}[P_{N}(f_{k'})]}}.
\end{equation*}
The numerator quickly drops to zero after a few bins (as in \cref{fig:correlation-vs-freq}), so we can restrict the computation to close bins: $|k-k'| \sim O(10)$. If the PSD is sufficiently smooth, we have $\operatorname{Var}\left[P_N(f_k)\right] \approx \operatorname{Var}\left[P_N(f_{k'})\right]  \approx S_{xx}(f_k)^2.$
Considering the numerator, we follow the same rationale as in \cref{sec:time-correlations}:
\begin{align}
	\label{eq:frequency-periodogram-covariance}
	\operatorname{Cov}\left[P_N(f_k), P_N(f_{k'})\right]  = \left(\frac{2}{\tau_s \kappa_N}\right)^2 \Big( & \left| \operatorname{E} [\tilde{x}_k \tilde{x}^{\ast}_{k'} ] \right|^2 \nonumber \\
	& + \left| \operatorname{E} [\tilde{x}_k \tilde{x}_{k'} ]\right|^2 \Big) .
\end{align}
The first expectation term in the parenthesis is directly given by the \gls{dft} covariance in  \cref{eq:frequency-covariance-elements}. The second term is similar to \cref{eq:second-term}, without the complex exponential. The window \gls{dft} being sharply peaked around zero, we can neglect the second term as in \cref{sec:time-correlations}, and we get
\begin{align}
    \operatorname{E} [\tilde{x}_k \tilde{x}^{\ast}_{k'} ] & \approx \frac{1}{2} S_{xx}(f_k) \int_{-\frac{f_s}{2}}^{+\frac{f_s}{2}} \tilde{w}_N(f_k - f)  \tilde{w}_N^{\ast}(f_{k'}- f) df.
\end{align}
We use the expression for the \gls{dft} of the window to get
\begin{equation}
\label{eq:periodogram-covariance-approx}
    \operatorname{Cov}\left[P_N(f_k), P_N(f_{k'})\right] \approx   \frac{S_{xx}(f_k)^2}{\kappa_N^2}\left| \sum_{n=0}^{N-1} w_n^2 e^{-\frac{2i \pi (k-k')}{N}}\right|^2.
\end{equation}
With this approximation, the correlation coefficient only depends on the window:
\begin{equation}
    \sigma_{kk'} \approx \frac{1}{\kappa_N^2}\left| \sum_{n=0}^{N-1} w_n^2 e^{-\frac{2i \pi (k-k')}{N}}\right|^2.
\end{equation}
Using this result in \cref{eq:effective-ndof-freq} and following the same rationale as in \cref{sec:time-correlations} leads to a simpler computation of the effective number of \glspl{dof}:
\begin{equation}
\label{eq:effective-ndof-freq-2}
    \nu(f_j) = \frac{2 N_j}{1+2\sum_{k=1}^{N_j-1} \left(1 - \frac{k}{N_j}\right) \gamma(k)},
\end{equation}
where we defined 
\begin{equation}
    \gamma(k) \equiv \frac{1}{\kappa_N^2}\left| \sum_{n=0}^{N-1} w_n^2 e^{-\frac{2i \pi k}{N}}\right|^2.
\end{equation}

\vspace{2cm}

\section{\label{sec:bias-proof}Estimation of the bias when using inaccurate PSD templates}

We derive the bias on the parameter vector $\boldsymbol{\theta}$ when using a \gls{psd} model $S(f, \boldsymbol{\theta})$ that inaccurately describes the actual statistical expectation of the periodogram.

We start from the likelihood in \cref{eq:averaged-loglike}. The best-fit parameter $\boldsymbol{\theta}_\mathrm{bf}$ verifies \cref{eq:best-fit-equation}, which yields
\begin{equation}
\label{eq:gradient-equals-zero}
    \sum_{j=j_{\min}}^{j_{\max}} \frac{1}{S(f_j, \boldsymbol{\theta}_\mathrm{bf})^2} \frac{\partial S(f_j)}{\partial \theta_l} \left(\bar{P}_L(f_j) - S(f_j, \boldsymbol{\theta}_\mathrm{bf})\right)=0.
\end{equation}

At first order in the error, we have
\begin{equation}
\label{eq:psd-error}
    S(f, \boldsymbol{\theta}_{\mathrm{tr}}) - S(f, \boldsymbol{\theta}_{\mathrm{bf}}) = - \boldsymbol{\nabla} S(\boldsymbol{\theta}_{\mathrm{bf}}) ^\intercal \Delta \boldsymbol{\theta},
\end{equation}
where $\boldsymbol{\nabla} S$ is the gradient of the \gls{psd} with respect to $\boldsymbol{\theta}$.
Plugging \cref{eq:psd-error} into \cref{eq:gradient-equals-zero} yields
\begin{align}
\label{eq:gradient-equals-zero-2}
    \frac{\nu}{2}\sum_{j=j_{\min}}^{j_{\max}} & \frac{1}{S(f_j, \boldsymbol{\theta}_\mathrm{bf})^2} \frac{\partial S(f_j)}{\partial \theta_l} \nonumber \\
    & \left( \bar{P}_L(f_j) - S(f_j, \boldsymbol{\theta}_\mathrm{tr}) -  \boldsymbol{\nabla} S(\boldsymbol{\theta}_{\mathrm{bf}})^\intercal\Delta \boldsymbol{\theta}\right)=0.
\end{align}
We are interested in the average error, which is the bias without the effect of the statistical uncertainty. To that end, we take the expectation of \cref{eq:gradient-equals-zero-2},
which amounts to replacing $\bar{P}_L(f_j)$ by its expectation $\operatorname{E}[\bar{P}_L(f_j) \, |\,  \boldsymbol{\theta}_{\mathrm{tr}}]$ in the equation. We get
\begin{align}
\label{eq:gradient-equals-zero-3}
    & \frac{\nu}{2}\sum_{j=j_{\min}}^{j_{\max}}  \frac{1}{S(f_j, \boldsymbol{\theta}_\mathrm{bf})^2} \frac{\partial S(f_j)}{\partial \theta_l}\left( \operatorname{E}[\bar{P}_L(f_j) \, |\,  \boldsymbol{\theta}_{\mathrm{tr}}] - S(f_j, \boldsymbol{\theta}_\mathrm{tr})\right) \nonumber \\
   & = \frac{\nu}{2}\sum_{j=j_{\min}}^{j_{\max}} \frac{1}{S(f_j, \boldsymbol{\theta}_\mathrm{bf})^2} \frac{\partial S(f_j)}{\partial \theta_l} \boldsymbol{\nabla}S(\boldsymbol{\theta}_{\mathrm{bf}})^\intercal\Delta \boldsymbol{\theta} \nonumber \\
   & = \sum_{m=1}^{p}  \frac{\nu}{2}\sum_{j=j_{\min}}^{j_{\max}} \frac{1}{S(f_j, \boldsymbol{\theta}_\mathrm{bf})^2} \frac{\partial S(f_j)}{\partial \theta_l} \frac{\partial S(\boldsymbol{\theta}_\mathrm{bf})}{\partial \theta_m} \Delta \theta_m \nonumber \\
   & = \mathbf{I}(\boldsymbol{\theta}_{\mathrm{bf}}) \Delta \boldsymbol{\theta}
\end{align}
Inverting this relation yields
\begin{equation}
     \Delta \boldsymbol{\theta} = \mathbf{I}(\boldsymbol{\theta}_{\mathrm{bf}})^{-1} \mathbf{B},
\end{equation}
where $\mathbf{B}$ is the $p-$column vector defined by
\begin{equation}
    B_{l} \equiv \frac{\nu}{2}\sum_{j=j_{\min}}^{j_{\max}}  \frac{1}{S(f_j, \boldsymbol{\theta}_\mathrm{bf})^2} \frac{\partial S(f_j)}{\partial \theta_l} \operatorname{Bias}[\bar{P}_L(f_j) \, | \, \boldsymbol{\theta}_\mathrm{tr}].
\end{equation}
In the expression above, we use the spectrum bias as in \cref{eq:spectrum-bias}, i.e., the difference between the periodogram expectation and the target \gls{psd}:
\begin{equation}
  \operatorname{Bias}[\bar{P}_L(f_j) \, | \, \boldsymbol{\theta}_\mathrm{tr}] = \operatorname{E}[\bar{P}_L(f_j) \, |\,  \boldsymbol{\theta}_{\mathrm{tr}}]  - S(f_j, \boldsymbol{\theta}_\mathrm{tr}).
\end{equation}

\vspace{2cm}

\bibliographystyle{apsrev4-2-short}
\bibliography{references}

\end{document}